\documentclass[floatfix,
        amsfonts,
        aps, 
        prd, 
        12pt,
        eqsecnum,
        bibtotoc,
        tightenlines,
        superscriptaddress,
        preprintnumbers,
        nofootinbib
]{revtex4}

\usepackage{graphicx}
\usepackage{amsmath,amssymb,amstext,xcolor}
\usepackage{bm}                         

\newcommand{\OF}{{\mathcal O}_{F^2}}

\newcommand{\wn}{\mathfrak{w}}
\newcommand{\qn}{\mathfrak{q}}

\newcommand{\q}{{\bf q}}
\def\k{{\bf k}}
\newcommand{\dd}{\mathrm{d}}

\newcommand{\tr}{\mathop{\mathrm{tr}}}
\renewcommand{\Im}{\mathrm{Im}}

\renewcommand{\L}{\mathcal{L}}

\newcommand{\threeVec}[1]{\ensuremath{\mathrm{\bf #1}}}

\newcommand{\M}{{\mathcal M}}
\newcommand{\x}{{\bf x}}
\newcommand{\p}{{\bf p}}
\newcommand{\pbar}{{\bar{p}}}
\newcommand{\llangle}{\left\langle} 
\newcommand{\rrangle}{\right\rangle}
\newcommand{\E}{\mathcal{E}}
\newcommand{\F}{\mathcal{F}}
\newcommand{\N}{\mathcal{N}}

\renewcommand{\O}{\mathcal{O}}
\def\Eq#1{Eq.~(\ref{#1})}
\def\st{\begin{equation}}
\def\stp{\end{equation}} 
\def\bg{\begin{eqnarray}}
\def\nd{\end{eqnarray}}
\def\App#1{Appendix~\ref{#1}}
\def\Fig#1{Fig.~\ref{#1}}
\def\Sect#1{Section~\ref{#1}}
\def\Ref#1{Ref.~\cite{#1}}

\begin{document}

\preprint{MPP-2008-46}

\title{Quarkonium transport in thermal AdS/CFT}

\author{Kevin Dusling}
        \email{kdusling@gmail.com}
        \affiliation{Department of Physics and Astronomy, Stony Brook University , Stony Brook NY 11794-3800, USA}
\author{Johanna Erdmenger}
        \email{jke@mppmu.mpg.de}
        \affiliation{Max-Planck-Institut f\"ur Physik (Werner-Heisenberg-Institut), F\"ohringer Ring 6, 80805 M\"unchen, Germany}
\author{Matthias Kaminski}
        \email{kaminski@mppmu.mpg.de}
        \affiliation{Max-Planck-Institut f\"ur Physik (Werner-Heisenberg-Institut), F\"ohringer Ring 6, 80805 M\"unchen, Germany}
\author{Felix Rust}        
        \email{rust@mppmu.mpg.de}
        \affiliation{Max-Planck-Institut f\"ur Physik (Werner-Heisenberg-Institut), F\"ohringer Ring 6, 80805 M\"unchen, Germany}
\author{Derek Teaney}
    \email{derek.teaney@stonybrook.edu}
        \affiliation{Department of Physics and Astronomy, Stony Brook University , Stony Brook NY 11794-3800, USA}
\author{Clint Young}
        \email{young@tonic.physics.sunysb.edu}
        \affiliation{Department of Physics and Astronomy, Stony Brook University , Stony Brook NY 11794-3800, USA}

\begin{abstract}
\vspace{1em}
\noindent
We consider a heavy meson moving slowly through  high temperature
non-abelian plasmas.  Using a simple dipole effective Lagrangian, we
calculate the in-medium mass shift  and the drag coefficient  of the
meson  in $\N=4$ Super Yang Mills theory at weak and strong coupling.
As anticipated, in the large $N$ limit
the mass shift is finite while  the drag  is suppressed by $1/N^2$ .
After comparing
results to perturbative QCD estimates (which are also calculated), we
reach the conclusion that relative to weak coupling expectations 
the effect of strong coupling is 
to reduce the momentum diffusion rate and to {\it increase} the relaxation time by up to a factor of four. 

\end{abstract}


\maketitle


\section{Introduction} 
\label{sec:introduction}

 The energy loss of heavy quarks  and quarkonia in media has been a
subject of intense experimental interest
\cite{Adare:2006nq,Bielcik:2005wu,Adare:2008sh,Adare:2006ns,Adler:2005ph,Arnaldi:2006ee,Alessandro:2004ap}. The suppression of charm and bottom quarks
observed at RHIC  motivated several groups to utilize the 
guage-gravity duality \cite{Maldacena:1997re,Witten:1998qj,Gubser:1998bc,Aharony:1999ti} 
to compute the 
drag of fundamental heavy quarks in $\N=4$ Super Yang-Mills 
at strong coupling 
\cite{Herzog:2006gh,CasalderreySolana:2006rq,Gubser:2006bz}. 
The goal of this paper is to extend
these strong coupling calculations of drag and diffusion from heavy quarks to heavy mesons. 

The motivation for this effort is twofold. First, future experiments at
RHIC promise to measure the elliptic flow of $J/\psi$ mesons, and it is
important to support this experimental program with theoretical work.
To this end, various groups have studied the thermal properties of
heavy mesons  within the context of  the AdS/CFT correspondence \cite{Liu:2006nn,Peeters:2006iu,Mateos:2007vn,Ejaz:2007hg}.  However,
in spite of this progress, the transport properties  of these mesonic
excitations are not well understood.  Although the kinetics derived in
this paper are not directly applicable to the heavy ion experiments, we
believe that the results do hold some important information for
phenomenology.

The second motivation for this work is theoretical. After the quark 
drag was computed using the correspondence,
it was realized that
the drag of quarkonia is zero in a large $N_c$ limit 
\cite{Liu:2006nn,Peeters:2006iu,Mateos:2007vn}.  
In the high temperature phase (when the temperature is comparable to the mass of lowest meson state)
the vector spectral function has been computed and it shows a rich dynamical picture of meson melting 
\cite{Hoyos:2006gb,Myers:2007we,Erdmenger:2007ja,Myers:2008cj,Erdmenger:2008yj}. 
However, in the low temperature phase describing heavy mesons, 
the spectral function is usually described by a sequence of states with 
zero width.  
Since within 
a thermal environment the 
drag and diffusion of these mesonic states is certainly not zero, 
it remained as a theoretical challenge to compute the kinetics
of these states  using the AdS/CFT setup. 

Rising to this challenge, the width of AdS/CFT mesons due to scattering
with surrounding heavy quarks (or anti-quarks) was  recently determined
by extending the analysis of meson melting to finite baryon density
\cite{Myers:2008cj} and by  studying string worldsheet instantons
at zero baryon density \cite{Faulkner:2008qk}.  In general the meson width determined in this way
is suppressed by the  density of heavy quarks. (At zero baryon density the
width is suppressed by the thermal population of heavy quarks.)   In contrast, we are
concerned with the thermal width which  is finite at zero density and
infinite $\lambda$ and  captures the rescattering between the meson and
the surrounding $\N=4$ medium. 

This work will focus on heavy mesons where the binding energy is much
greater than the temperature.  While in perturbation theory the
constraint on the binding energy reads $m_q v^2 \gg T$,  in the
strongly coupled $\N=4$ theory the constraint is $\frac{2\pi
m_{q}}{\sqrt{\lambda}} \gg T$.  In this tight binding regime, mesons
survive well above $T_c$ and the meson width is sufficiently narrow to
speak sensibly about drag and momentum diffusion.

For real charmonium (bottomonium) the binding energy 
can be estimated from  mass splitting
between the $2s$ and $1s$ ($3s$ and $1s$) states,  
$\Delta M_{2s-1s}^{J/\psi} \simeq 589 \,{\rm MeV} $  and $\Delta M_{3s-1s}^{\Upsilon} \simeq 895 \, {\rm MeV}  $ respectively \cite{Yao:2006px}.  Therefore it is not really clear that
real quarkonia above $T_c \simeq 170\leftrightarrow 190\,{\rm MeV}$
\,\cite{Aoki:2006br,Cheng:2006qk} can be  modeled as a simple dipole
which 
lives long enough to be considered a quasi-particle. 
Indeed  weak coupling hot QCD calculations 
of the spectral function show 
that over the temperature range $g^2M  \leftrightarrow   gM$, the 
concept of a meson quasi-particle slowly transforms  from  being  well
defined  to being increasingly vague \cite{Laine:2006ns,Laine:2007gj,Laine:2007qy,Burnier:2007qm,Brambilla:2008cx}. 
There is lattice evidence 
based on the maximal entropy method (which is not without uncertainty) 
that $J/\psi$ and $\Upsilon$  survive  to $1.6\,T_c$ and $\sim\!3\,T_c$  respectively \cite{Umeda:2002vr,Asakawa:2003re,Datta:2003ww,Iida:2006mv,Jakovac:2006sf,Aarts:2007pk}. 
However, model potential  calculations  which fit all the Euclidean lattice correlators indicate that the the $J/\psi$ and $\Upsilon$ survive
only up to at most $1.2\,T_c$ and $2.0\,T_c$ respectively 
\cite{Mocsy:2007yj,Mocsy:2007jz}.  
Clearly, the word ``survive" in this context is qualitative and means that there is a
discernible  peak in the spectral function. 
Given 
these facts, the 
assessment of the authors is that the dipole approximation might be reasonable
for $\Upsilon(1S)$ but poor for charmonium states and other bottomonium 
states.

Within the context of guage gravity duality, mesons are  studied 
by exploiting convenient generalizations of the AdS/CFT
correspondence in which fundamental flavor degrees of freedom 
are added by inserting additional probe branes into the geometry. In this paper we will use the approach due to Karch
and Katz \cite{Karch:2002sh} in which the additional quark flavors
are obtained by
adding D7 brane probes to the original D3 brane setup. On the gravity side,
the probe branes wrap a subspace which asymptotically near the boundary is
$AdS_5 \times S^3$. The meson spectrum for the dual $\N=2$ supersymmetric
gauge theory was first calculated in \cite{Kruczenski:2003be} by considering
fluctuations of the D7 branes embedded. Restricting this result to
fluctuations with vanishing angular momentum on the $S^3$, the meson spectrum
is given by 
\begin{equation} \label{susymass}
M =  \frac{2\pi m_q}{ \sqrt{\lambda}} \, 2 \sqrt{(n+1)(n+2)} 
\; ,
\end{equation}
with $m_q$ the quark mass determined by the separation between the D3 and D7
branes, $\lambda$ the 't~Hooft coupling and $n$ the radial excitation number. 
Subsequently, a gravity dual of chiral symmetry breaking has been obtained in
\cite{Babington:2003vm} by embedding a D7 brane probe into a deformed
non-supersymmetric gravity background with non-trivial dilaton
\cite{Constable:1999ch}. In this case there is a Goldstone boson
in the meson spectrum. The thermodynamics of mesons has been studied within
gauge/gravity duality by embedding a D7 brane probe into the AdS-Schwarzschild
black hole background. Within the deconfined phase, there is a new fundamental
first order phase transition which corresponds to meson melting
\cite{Babington:2003vm, Kirsch:2004km, Kruczenski:2003uq, Mateos:2006nu, Hoyos:2006gb}. 
For a review on mesons in the AdS/CFT correspondence see 
\cite{Erdmenger:2007cm}.

In this paper we consider a heavy meson moving slowly through the medium.  We
perform both a perturbative QCD and a strong coupling $\N=4$ SYM computation. 
For both approaches we first calculate the in medium meson mass shift, which 
determines the polarizabilities of the meson. 
As expected from the dipole effective theory,   the mass shift scales as
$T^4/\Lambda_B^3$, with $T$ the temperature and $\Lambda_B$ the inverse size of the meson. In the perturbative
calculation, $\Lambda_B$  is the inverse Bohr radius, while in
the AdS/CFT computation the meson mass plays this role.
In the $\N=4$  field theory the dipole effective Lagrangian  couples the heavy meson 
to the stress tensor and the square of the field strength
$\O_{F^2}$. In AdS/CFT we obtain these couplings from the linear response
of the meson mass to switching on a black hole background
or a non-trivial dilaton flow, respectively.
For the dilaton
flow we consider the $D3+D(-1)$ gravity background of Liu and Tseytlin
\cite{Liu:1999fc}. This background and the AdS-Schwarzschild background 
 allow for an analytic calculation of the meson polarizabilities.

Using these polarizabilities we subsequently calculate  
the momentum broadening $\kappa$ 
and the drag coefficient $\eta_{\scriptscriptstyle D}$.
This requires the calculation of
two-point functions involving gradients of the stress tensor and the field strength
squared. Within gauge/gravity duality, these are  obtained by considering
graviton and dilaton propagation through the AdS-Schwarzschild black hole
background. 


An outline of the paper is as follows. First, in \Sect{sec:perturbative}
we review the computation of drag and diffusion of
heavy $Q\bar{Q}$ bound states within the setup of pertubative
QCD. This will outline a two step procedure to determine the
drag coefficient at strong coupling. 
The first step is to 
determine the in medium mass shift (which is finite at large $N_c$) which determines the polarizabilities of the meson. This is 
done in section \Sect{sec:finiteT}. The second step  is to compute
the force-force correlator on the meson using the previously computed polarizabilities. This determines the drag and diffusion coefficient 
as reviewed in \Sect{sec:adsCftCorrelators}. Finally we compare 
our results to perturbation theory and reach some conclusions for
the RHIC experiments in \Sect{sec:summary}.

\section{Diffusion of Heavy Mesons in Perturbative large $N$ field theories } 
\label{sec:perturbative}

\subsection{Diffusion of Heavy Mesons in Perturbative large $N$ QCD } 
The interactions of a heavy meson with the QCD medium is 
well described by a dipole approximation. This physical approximation
has been formalized in the language of heavy meson effective 
Lagrangrians which we will adopt \cite{Luke:1992tm}. This perturbative
scheme relies on the large mass of the meson relative to the
external momenta of the gauge fields  
but does not rely on the smallness of
the coupling constant. It was used previously to make a good estimate
for the binding of $J/\psi$ to nuclei \cite{Luke:1992tm}.  

The heavy meson field $\phi_v$ describes a (scalar) meson which has
a fixed velocity $v^{\mu} = (\gamma,\gamma {\bf v})$. 
Then the effective Lagrangian for this meson field interacting with the
gauge fields is 
\begin{equation}  
\label{eq:leff}
 \L_{\rm eff} = - \phi^{\dagger}_v i v\cdot\partial \phi_v  +  
\frac{c_{E}}{N^2} \phi_v^\dagger \O_E \phi_v  + 
\frac{c_{B}}{N^2} \phi_v^\dagger \O_B \phi_v 
\; ,
\end{equation}  
with
\begin{gather}
\O_E = -\frac{1}{2} G^{\mu \alpha A}G_{\alpha}{}^{\nu A}  v_\mu v_\nu 
\; , 
\quad 
\O_B = \frac{1}{4}  G^{\alpha \beta A} G_{\alpha
  \beta}{}^A  - \frac{1}{2} G^{\mu \alpha A}G_{\alpha}{}^{\nu A}  v_\mu v_\nu 
\; . 
\end{gather}
$G^{\mu\nu}$ is the non-Abelian field strength of QCD, and $c_E$ and $c_B$ are
matching coefficients (polarizabilities) to be determined from the QCD dynamics
of the heavy $Q\bar{Q}$ pair. In inserting a factor of $1/N^2$ into the
effective Lagrangrian we have anticipated that the couplings of the heavy meson
to the field strengths are suppressed by $N^2$ in the large $N$ limit.

In the rest frame of a heavy quark bound state, $v=(1,0,0,0)$, the
operators $\O_E$ and $\O_B$ are
\begin{gather}
\O_E = \frac{1}{2} \,{\bf E}^A \cdot {\bf E}^A 
\; , \qquad 
\O_B = \frac{1}{2} \, {\bf B}^A \cdot {\bf B}^A 
\; ,
\end{gather}
where ${\bf E}^A$ and ${\bf B}^A$ are the color electric and magnetic fields. If
the constituents of the dipole are non-relativistic it is expected that the
magnetic polarizability $c_{B}$ is $O(v^2)$ relative to the electric
polarizability. For heavy quarks (where $c_B$ is neglected) and large $N$ these
matching coefficients were computed by Peskin \cite{Peskin:1979va,Bhanot:1979vb}
\begin{equation}    
\label{eqn:peskin}
  c_{E} = \, \frac{28\pi}{3\Lambda_B^3} 
\; , 
\qquad c_B=0 
\; .
\end{equation}
Here $\Lambda_B\equiv 1/a_0 = (m_q/2) C_F \alpha_s$ is the inverse Bohr radius for a $Q\bar{Q}$ bound state. It
 is finite at large $N$ since with $C_F\simeq N/2$ and finite $\lambda$ we have
$\Lambda_{B}=m_q \lambda /{16\pi}$. We will assume that
$c_B=0$ is zero in this section and subsequently generalize our results to $\N=4
$ theory.

This effective Lagrangian can be used to 
compute both the thermodynamics
and kinetics of the heavy meson state. 
First, to evaluate the in medium mass shift 
one
simply uses first order perturbation theory 
$\delta M = 
\llangle H_{\rm I} \rrangle =-\llangle \L_{\rm I} \rrangle $, yielding
\begin{equation}
\label{eq:qcd_massshift}
\begin{aligned}
    \delta M &= -\frac{c_{E}}{N^2} \llangle \O_E \rrangle_T \\
             &= -T \left(\frac{\pi T}{\Lambda_B}\right)^3\, \frac{14}{45}
 \; .
\end{aligned}
\end{equation}  
In the second line we have simply calculated the  expectation 
value
$\llangle \O_{E} \rrangle_{T} = \frac{\pi^2}{30} N^2 T^4 $ in a free 
gluon gas and used \Eq{eqn:peskin}. 

The importance of this result is that it is finite at large $N$ and that it is
in general suppressed by $(T/\Lambda_B)^3$, i.e. suppressed by powers of the
hadron scale to the temperature. If higher dimension operators were added to the
effective Lagrangian their contributions would be suppressed by additional
powers of $T/\Lambda_B$. At strong coupling, we will use the AdS/CFT
correspondence to determine the $\N=4$ polarizabilities from the mass shift.

We turn next to the kinetics of a heavy dipole in the medium.  
For timescales which are long compared to medium correlations,  
we expect that the kinetics of the 
heavy meson is described by  Langevin equations 
\begin{eqnarray}
\label{newton_langevin}
\frac{\dd p_i}{\dd t} &=& \xi_i(t) - \eta_{\scriptscriptstyle D} p_i 
\; , 
\qquad \langle \xi_i(t) \xi_j(t') \rangle = \kappa \delta_{ij}
        \delta(t-t') 
\; .
\end{eqnarray}
Here $\xi_i$ is a random force with second moment $\kappa$ and 
$\eta_{\scriptscriptstyle D}$ is the drag coefficient.
The drag and fluctuation coefficients are related  through the 
Einstein relation 
\begin{equation}
\label{etad}
\eta_{\scriptscriptstyle D} = \frac{\kappa}{2 MT} 
\; .
\end{equation} 
The Langevin equation is valid for times which are long compared to the 
inverse  temperature but short compared to the lifetime of the  
quasi-particle state. 

We can use the effective dipole Lagrangian  to calibrate the noise  
of the stochastic evolution, $\kappa$. 
The  microscopic equations of motion 
for a heavy particle in the medium are 
\begin{eqnarray}
   \frac{\dd p^i}{\dd t} = \mathcal{F}^i(t) 
\; ,
\end{eqnarray}
where $\F$ is a phenomenological force to be specified  below.
We then compare the response of the Langevin process 
to  the microscopic theory. 
Over a time which is long compared to medium correlations but 
short compared to the time scale of equilibration we can neglect the 
drag and equate the stochastic process to the microscopic theory  
\begin{equation}
   \int \! \dd t \int \! \dd t' \, \llangle \xi_i(t) \xi_j(t') \rrangle = \mbox{(time)}  \times  \kappa \, \delta_{ij} = 
    \int \! \dd t  \int \! \dd t' \,\llangle \mathcal F_i(t) \mathcal F_j(t') \rrangle   
\; .
\end{equation}
In a rotationally invariant medium we have
\begin{eqnarray}
\label{ff}
    \kappa & = &  \frac{1}{3} \int \!\dd t \, \llangle \F^i(t) \F^i(0)  \rrangle 
\; .
\end{eqnarray}
In the present context we identify the force with the 
negative of the gradient of the interaction Hamiltonian $H_{I} = -\L_I$ ,
\begin{equation}
 {\bm{\mathcal F}(t)} = \int\! \dd^3\x\, 
\phi_v^\dagger(\x,t) \, \left[c_E \bm{\nabla} \mathcal{O}_E(\x,t) \right] \phi_v(\x,t)
\; ,
\end{equation}
which is the usual form of a dipole force averaged over the 
wave function of the meson.
The fluctuation dissipation theorem  relates the correlation function 
in \Eq{ff} (with the specified time order of operators) 
to the imaginary part of the retarded force-force correlator
\begin{equation} 
\label{ff2}
    \kappa =  \frac{1}{3}\; \lim_{\omega\rightarrow 0}  \frac{-2T}{\omega} \Im G_{R}(\omega)
\; ,
\end{equation} 
where the retarded correlator  is
\begin{equation}  
  G_{R} = -i\int\! \dd t \, e^{+i\omega t} \,  \theta(t) \llangle \left[ \F^i(t), \F^i(0) \right] \rrangle
\; .
\end{equation}  
Integrating out the heavy meson field as discussed in detail in  
\Ref{CasalderreySolana:2006rq}, which treated the heavy quark case, 
we obtain a formula for the momentum diffusion coefficient 
\begin{equation} 
\label{e2e2formula}
  \kappa = \frac{1}{3} \frac{c_E^2}{N^4}
\int \! \frac{\dd^3\q}{(2\pi)^3}\,\q^2\left[-\frac{2 T}{\omega}  \Im G_{R}^{\mathcal O_E\mathcal O_E} (\omega,\q) \right]  
\; ,
\end{equation}  
with the retarded $\mathcal O_E \mathcal O_E$ correlator given by
\begin{equation}  
 G_{R}^{\mathcal O_E\mathcal O_E}(\omega,\q) = -i\int \! \dd^4x \, e^{+i\omega t - i\q\cdot \x} 
\theta(t)\llangle \left[\mathcal O_E(\x,t)\, ,\, \mathcal O_E({\bf 0},0) \right] \rrangle 
\; .
\end{equation}  

\begin{figure}
\begin{center}
\includegraphics[width=2.5in]{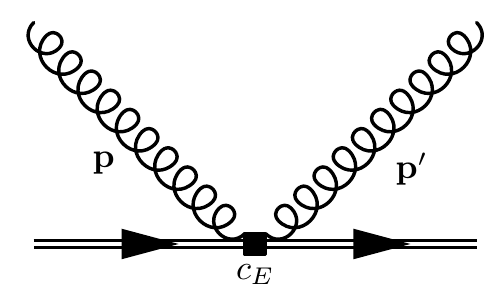}
\end{center}
\caption{ 
\label{graph1}
Dipole scattering graph which causes drag and diffusion of heavy mesons 
in QCD and $\N=4$ SYM. The QCD correlator which encodes this physics is 
given by \Eq{e2e2formula}. }
\end{figure}

We can understand this result with simple kinetic theory. Examining the Langevin
dynamics we see that $3\kappa$ is the mean squared momentum transfer to the
meson per unit time. The factor of three arises from the number of spatial
dimensions. In perturbation theory this rate is easily computed by weighting the
transition rate for any gluon in the bath to scatter with the heavy quark by the
square of the momentum transfer,
\begin{equation}  
\label{eq:kappaqcd}
 3\kappa =  \int \! \frac{\dd^3\p}{(2\pi)^32E_\p}\,\frac{\dd^3\p'}{(2\pi)^3 2E_{\p'}} \, \left|\M\right|^2\, n_\p (1 + n_{\p'}) \, \q^2\, (2\pi)^3 \delta^{3}(\q - \p + \p') 
\; .
\end{equation}  
Here $\p$ is the incoming gluon, $\p'$ is the outgoing gluon 
and $\q$ is the momentum transfer $\q=\p-\p'$ as indicated in 
\Fig{graph1}.
$\left|\M \right|^2$ 
is the gluon meson scattering amplitude computed with the effective 
Lagrangian in \Eq{eq:leff}  and summed over colors and helicities of the incoming and 
outgoing gluon
\begin{equation}  
  \left|\M \right|^2 =  \frac{c_{E}^2}{N^2}  \omega^4 \,\left(1 + \cos^2(\theta_{\p\p'})\right)
\; .
\end{equation}  
Alternatively (as detailed in \App{perturb})
we can simply evaluate the imaginary part of the 
retarded amplitude written in \Eq{e2e2formula}  
to obtain the same result. 

For QCD the integrals written in 
\Eq{eq:kappaqcd} are straightforward 
and yield the following result for the rate of momentum broadening 
\begin{equation} 
\label{eq:kappaqcd_final}
\begin{aligned}
      \kappa &= \frac{1}{N^2} c_E^2 \frac{64 \pi^5}{135} T^9  \\
             &= \frac{T^3}{N^2} \left( \frac{\pi T}{\Lambda_B} \right)^6\frac{50176\pi}{1215} 
\; .
\end{aligned}
\end{equation}  
The high power of temperature $T^9$ arises since the dipole cross section
rises as $\omega^4$.  
The matching coefficient $c_E$ is directly related to the mass shift of the dipole and encodes the coupling of the long distance gluonic fields to the dipole. By taking 
the ratio between the momentum broadening and the mass shift squared we 
find a physical quantity which is independent of this  coupling
\begin{equation}  
  \frac{\kappa}{(\delta M)^2}=  \frac{\pi T}{N^2} \, \frac{1280}{3}
\; .
\end{equation}  
The large numerical factor $1280/3$ originates from the cross section 
which grows  as $\omega^4$.  A similarly large factor  appears in
$\N=4$ SYM as discussed below in greater detail.

\subsection{Linear perturbations of $\N=4$ Super Yang-Mills theory}

Our aim is to calculate the heavy meson diffusion coefficient $\kappa$ from
gauge/gravity duality. This requires the calculation of the two-point
correlators as well as of the associated polarizabilities in $\N=4$ Super
Yang-Mills theory.

The same formalism used in the preceding section can be used for $\N=4$ $SU(N)$
Super Yang-Mills theory. In general all operators in $\N=4$ which are scalars
under under Lorentz transformations and $SU(4)$ R-charge rotations will couple
to the meson at some order. The contribution of higher dimensional operators is
suppressed by powers of the temperature to the inverse size of the meson. The
lowest dimension operator which could couple to the heavy meson field is
$\O_{X^2} = \mathop{\mathrm{tr}} X^{i} X^{i}$, where $X^i$ denotes the scalar
fields of $\N=4$ theory. However the anomalous dimension of this operator is not
protected, and the prediction of the supergravity description of $\N=4$ SYM is
that these operators decouple in a strong coupling limit \cite{Aharony:1999ti}.
The lowest dimension gauge invariant local operators which are singlets under
$SU(4)$ and which have protected anomalous dimension are the stress tensor
$T_{\mu\nu}$ which couples to the graviton, minus the Lagrangian $\O_{F^2}
=-\L_{N=4}$, which couples to the dilaton\footnote{Since we can add a total
derivative to the Lagrangian, the operator $-\L$ is ambiguous. The precise form
of the operator coupling to the dilaton is given in \Ref{Klebanov:1997kc}. We
neglect this ambiguity here.} and the operator $\O_{F\tilde F} =
\mathop{\mathrm{tr}}F^{\mu\nu}{\tilde F}_{\mu\nu} + \ldots$, which couples to
the axion. An interaction involving $\O_{F\tilde F}$ breaks $CP$ which is a
symmetry of the Lagrangian $\N=2$ hypermultiplet of the $\N=4$ SYM gauge theory.
Thus interactions involving $\O_{F\tilde F}$ can be neglected.

Summarizing the preceding discussion, we find that the effective Langrangian
describing the interactions of a heavy meson coupling to the operators in the
field theory is
\begin{equation}
\label{eq:LeffN4}
\L_{\rm eff} = -\phi_v(\x,t) iv\cdot\partial \phi_v(\x,t) \, + \, \frac{c_T}{N^2} \phi_v^{\dagger} (\x,t) T^{\mu\nu}v_{\mu}v_{\nu}  \phi_v(\x,t) \, + \, \frac{c_{F}}{N^2} \phi_v^\dagger(\x,t)  
\O_{F^2} \phi_v(\x,t)  
\; ,
\end{equation}
which is a linear perturbation of $\N=4$ Super Yang-Mills theory by two
composite operators.

The polarization coefficients $c_T$, $c_F$ will be determined below from meson
mass shifts in gauge/gravity duality. This requires breaking some of the
supersymmetry. For the contribution of the energy-momentum tensor, this is
achieved by switching on the temperature. Then, the mass shift of the meson is
given by expectation value of the stress tensor
\begin{equation}  
\label{eq:mass_finiteT}
     \delta M =  -\frac{ c_{T}}{N^2} \llangle T^{00} \rrangle 
\; ,
\end{equation}
In gauge/gravity duality
this is achieved by considering the AdS-Schwarzschild black hole background 
where $\llangle \OF \rrangle=0$.
In contrast, for the meson response to $\llangle \OF\rrangle$ we consider a background  self-dual gauge configuration 
where $\llangle \OF\rrangle \neq 0$ while $\llangle T^{\mu\nu}\rrangle=0$. As can
be seen from the supersymmetry transformations of the $\N=4$ fermions, such a background breaks the supersymmetry to $\N=2$. The mass
shift of a heavy meson  is then
\begin{equation}  
\label{eq:mass_finiteF}
     \delta M =  -\frac{ c_{F}}{N^2} \llangle \O_{F^2} \rrangle
\; ,
\end{equation}
where
$\llangle \O_{F^2} \rrangle = \llangle - {\cal L} \rrangle$ 
is the expectation value of  the $\N=4$ Lagrangian in  
the $\N=2$ symmetric background configuration. A
suitable dilaton solution corresponding to this selfdual configuration been given by
Liu and Tseytlin \cite{Liu:1999fc} and will be used below to determine $c_F$.

As explained above for the QCD case, to determine the kinetics of the heavy
meson in the $\N=4$ background we identify the force on the heavy meson as minus
the gradient of the interaction Hamiltonian $H_I=-\L_I$
 averaged over
the meson wave function. With \Eq{eq:LeffN4}  we have
\begin{equation}  
{\bm \F}(t) = \int\! \dd^3\x \,\phi_v^\dagger(\x,t)\,\nabla\left[\frac{c_T}{N^2} T^{\mu\nu}u_{\mu}u_{\nu}  + \frac{c_{F^2}}{N^2} \O_{F^2} \right] \phi_v(\x,t)
\; .
\end{equation}  
Then  integrating out the heavy meson fields as above,
the force-force correlator in \Eq{ff2} 
becomes 
\begin{gather} \label{eq:adsCftCorrelator}
\kappa = \lim_{\omega\rightarrow0} 
\int \!\frac{\dd^3\q}{(2\pi)^3}\, \frac{\q^2}{3} 
\left[\left(\frac{c_T}{N^2}\right)^2\, \frac{-2T}{\omega}\Im G_{R}^{TT}(\omega,\q) 
  + \left(\frac{c_F}{N^2}\right)^2\,\frac{-2T}{\omega}\Im G_{R}^{F^2F^2}(\omega,\q)
\right ] 
\; .  
\end{gather}
where the retarded correlators are 
\begin{eqnarray}
\label{eq:correlators}
G_{R}^{TT} &=& -i\int\! \dd^4x\, e^{+i\omega t - i \q\cdot \x} \theta(t)\llangle \left[T^{00}(\x,t) , T^{00}({\bf 0},0) \right] \rrangle
\; , \\
G_{R}^{FF} &=& -i\int\! \dd^4x\, e^{+i\omega t - i \q\cdot \x} \theta(t)
\llangle\left[ \O_{F^2}(\x,t) , \O_{F^2}({\bf 0}, 0) \right] \rrangle 
\; .
\end{eqnarray} 
In writing  \Eq{eq:adsCftCorrelator} we have implicitly assumed
that there is no cross term  between $\O_{F^2}$ and $T^{\mu\nu}v_{\mu}v_{\nu}$.
In
the gauge/gravity duality this is reflected in the fact that at tree
level in supergravity, 
$\sim \frac{\delta^2S_{SUGRA}}{\delta
g^{00} (x) \delta \Phi (y)}=0$.

In summary, we first will determine the polarizabilites $c_{F}$ and
$c_{T}$ from the mass shifts of the meson in two different backgrounds using
\Eq{eq:mass_finiteT} and \Eq{eq:mass_finiteF}. Subsequently we will compute the
correlators in \Eq{eq:correlators} for $\N=4$ theory 
at finite temperature. Finally we will put
the results together using \Eq{eq:adsCftCorrelator} to deduce the rate of
momentum broadening.

\section{Determining Matching Coefficients with Mass Shifts} 
\label{sec:finiteT}

\subsection{Backgrounds dual to finite temperature and field strength}

\subsubsection{Finite temperature background}

The gravity background dual to  $\N=4$ SYM theory at finite temperature is given
by the AdS-Schwarzschild black hole with Lorentzian signature (see
e.g.~\cite{Policastro:2002se}). 
This background is needed below
 both for calculating the necessary two-point correlators $\langle
T^{00} T^{00} \rangle$ and $\langle \O_{F^2} \O_{F^2} \rangle$, as well as for
obtaining the meson polarizability,  $c_{T}$.

We make use of the coordinates of \cite{Babington:2003vm} to write the
AdS-Schwarzschild background with Lorentzian signature as
\begin{equation}
\label{eq:adsBHMetric}
\dd s^2 = \frac{w^2}{R^2}\left(-\frac{f^2}{\tilde f}\,\dd t^2 + \tilde f \dd\threeVec x^2 \right)
                                + \frac{R^2}{w^2} \left( \dd \varrho^2 + \varrho^2 \dd\Omega_3^2 + \dd w_5^2 + \dd w_6^2 \right),
\end{equation}
with the metric $\mathrm d \Omega_3^2$ of the unit $3$-sphere, and
\begin{equation}
\begin{gathered}
\label{eq:metricDefinitions}
f(r)=1-\frac{r_H^4}{4w^{4}}, \qquad \tilde f(r)=1+\frac{r_H^4}{4w^{4}},\qquad  w^2 = \varrho^2 + w_5^2 + w_6^2, \\
r_H = T \pi R^2,\qquad R^4=4\pi g_s N  \ell_s^4, \qquad \lambda=4\pi g_s N,\qquad g_{YM}^2=g_s
\; .
\end{gathered}
\end{equation}
This spacetime has a horizon at $w_H$ which is determined by $r_H$ as
\begin{equation}
        w_H = \frac{r_H}{\sqrt 2},
\end{equation}
and the boundary is reached at asymptotically large $w$. With $r_H=0$ we
obtain $AdS_5\times S^5$.
In section~\ref{sec:adsCftCorrelators} we will work in a coordinate system with inverted 
radial AdS coordinate $u$ used in e.g.~\cite{Policastro:2002se}. In these
coordinates, the metric reads
\begin{equation} 
\dd s^2 = \frac{(\pi T R)^2}{u} 
  \left (-f(u)\, \dd t^2 +\dd \threeVec{x}^2 \right )+ 
  \frac{R^2}{4 u^2 f(u)}\,\dd u^2 + R^2 \dd\Omega_5^2 
\; ,     
\end{equation}
with~$f(u)=1-u^2$ and $\dd\Omega_5^2$ the unit $5$-sphere metric.

\subsubsection{Dilaton background}

A non-trivial dilaton background that is dual to a field configuration with $\llangle \OF \rrangle \neq 0$  and $\llangle T^{\mu\nu}\rrangle = 0$ 
has been given by Liu and Tseytlin \cite{Liu:1999fc} and  consists of a
configuration of D3 branes with homogeneously distributed D(-1) instantons. The
type IIB action in the Einstein frame for the dilaton~$\Phi$, the axion~$C$, and
the self-dual gauge field strength~$F_5=\star F_5$ reads
\begin{equation}
\label{eq:IIBAction}
S_{\scriptscriptstyle {\rm IIB}}=\frac{1}{2\kappa_{10}^2}\int\! \dd^{10}x\sqrt{-g}\; \left[
  \mathcal{R}-\frac{1}{2}(\partial \Phi)^2 -\frac{1}{2}e^{2\Phi}  
   (\partial C)^2 -  \frac{1}{4\cdot 5!} (F_5)^2 +\dots
\right]\: .
\end{equation}
The ten-dimensional Newtons constant is 
\begin{equation}  
        \label{eq:gravityConst10} 
        \frac{1}{2\kappa_{10}^2}=\frac{1}{(2\pi)^7 \ell_s^8 g_s^2}= 
        \frac{N^2}{4\pi^5 R^8}  \; .
\end{equation}
Solving the equations of motion derived from~\eqref{eq:IIBAction}, 
Liu and Tseytlin \cite{Liu:1999fc}  obtain the metric
\begin{equation}  
\label{eq:liuTseytlinBackground}
{\dd s}^2_{\text{string}} = 
e^{\Phi/2} \,{\dd s}^2_{\text{Einstein}}   
    = e^{\Phi/2}   
     \left[ 
      \left(\frac{r}{R}\right)^2 \eta_{\mu\nu}\dd x^\mu \dd x^\nu + 
      \left(\frac{R}{r}\right)^2 
      \left(\dd r^2 + r^2 \dd\Omega_5^2\right ) 
     \right]  
\; .
\end{equation}
 and axion-dilaton solution\footnote{For the
conventions note that in our notation~$q=\frac{R^8}{\lambda}q_{\text{LT}}$,
where $q_{LT}$ is used in the paper of Liu and Tseytlin \cite{Liu:1999fc}.}
\begin{equation}
        \label{eq:dilatonSolution}  
        e^\Phi= 1 + \frac{q}{r^4} \: ,  \qquad  C=-i\left(e^{-\Phi} - 1\right)\: .   
\end{equation} 
The expectation value $\left<\mathcal O_{F^2}\right>$ is given by 
\begin{equation} 
\label{eq:qandF2}
\langle \O_{F^2}(\vec{x}) \rangle = \lim_{r\rightarrow\infty} 
\frac{\delta S_{\scriptscriptstyle {\rm IIB}}}{\delta \Phi(r,\vec{x})}=  
  \frac{N^2}{2\pi^2 R^8}\, q 
\; .   
\end{equation}  

\subsection{Computing the polarization coefficients from meson mass shifts}

Heavy 
mesons are identified with fluctuations $\tilde \varphi$ of a D$7$ brane
embedded into the background dual to the field theory under consideration.
Stable embeddings are obtained if the D$7$ brane spans all Minkowski directions
as well as the radial AdS coordinate and a $3$-sphere in the remaining
polar directions. Consider the metric \eqref{eq:adsBHMetric} as an
example. The D$7$ brane spans all directions except $w_5$ and $w_6$. The meson
mass $M$ is then obtained by solving the equation of motion for the fluctuations
$\tilde \varphi$ \cite{Kruczenski:2003be}. Read as an eigenvalue equation, the
equation of motion gives the meson mass as the eigenvalues $M$ to the
corresponding eigenfunctions $\tilde \varphi$. The discrete values of $M$
describe the Kaluza-Klein mass spectrum of mesons for any given quark mass.

To see how this works, we outline this procedure for the vacuum case
$\left<T^{00}\right>=\left<\OF\right>=0$, for which the meson spectrum was
originally calculated in \cite{Kruczenski:2003be}. Subsequently we will
introduce a non-zero $\left<\OF \right>$ and $\left<T^{00}\right>$,
respectively.

In the case of a D$7$ brane embedded in a ten dimensional background, the brane
embedding is described by the location in the two directions transverse to the
brane. We call these directions $w_5$ and $w_6$. In general these locations 
depend on all eight coordinates $\xi^i$ of the eight dimensional D7 brane
worldvolume and are determined by extremizing the DBI-action
\begin{equation}
\label{eq:DBIaction}
        S_{\text{DBI}} = -T_7 \int\!\dd^8\xi \;e^{-\Phi} \sqrt{-\det h}\; , \qquad h_{ab} = 
\frac{\partial X^{\mu}}{\partial \xi^a} \frac{\partial X^{\nu}}{ \partial \xi^{b} }\,G^{\text{st}}_{\mu\nu} 
\; ,
\end{equation}
where $T_7$ is the D$7$ brane tension and $G^{\text{st}}$ is the string frame
metric of the ten dimensional background with coordinates $X^\mu$. It is related
to the Einstein metric as in \Eq{eq:liuTseytlinBackground}. The distinction
between the Einstein and string frame is ultimately important below. The
pullback $h$ contains the functions $w_5(\xi)$ and $w_6(\xi)$, which are
determined by solving their equations of motion, derived from $S_{\text{DBI}}$.

The background $AdS_5\times S^5$ dual to
$\left<T^{00}\right>=\left<\OF\right>=0$ is obtained e.g.\ from
\eqref{eq:adsBHMetric} with $r_H=0$. It is well known that for this background a
probe brane embedding is given by the functions
\begin{align}
        w_5 &= 0\: ,\\
        w_6 &= L \: ,
\end{align}
where $w_5$ and $w_6$ are the coordinates given in \eqref{eq:adsBHMetric} and
the constant $L$ determines the quark mass $m_q=L/(2\pi\ell_s^2)$. Now we allow
for small fluctuations $\tilde \varphi_5$ and $\tilde \varphi_6$ around this
solution,
\begin{align}
        w_5 &= 2\pi\ell_s^2 \:\tilde \varphi_5(\vec x,\varrho) \; ,\\
        w_6 &= L + 2\pi\ell_s^2\: \tilde\varphi_6(\vec x,\varrho) \; .
\end{align}
By the symmetries of the setup, the fluctuations only depend on the Minkowski
directions $\vec x$ and on the coordinate $\varrho$, denoting the radial
coordinate on the part of the D$7$ brane which is transverse to the Minkowski
directions. The resulting equations of motion are identical for $\tilde\varphi_5$ and
$\tilde\varphi_6$. Using $\varphi$ to denote any one of them, it was shown in
\cite{Kruczenski:2003be} that a solution to the equations of motion may be
obtained by separating variables with the ansatz
\begin{equation}
        \tilde\varphi = \varphi(\varrho)\,e^{i\,\vec k \vec x}\, \mathcal Y^\ell(S^3)\;,
\end{equation}
where $\mathcal Y^\ell (S^3)$ are the scalar spherical harmonics on the $S^3$
wrapped by the probe D$7$ brane and $\vec k$ denotes a four vector.  
The resulting equation of motion for the
radial function $\varphi(\varrho)$ may be obtained from \eqref{eq:DBIaction}. 
For $\ell=0$ it reads as
\begin{equation}
\label{eq:eomQ0}
        -\partial_\rho \rho^3 \partial_\rho \varphi(\rho) = \bar M^2\frac{\rho^3}{(\rho^2+1)^2}\,\varphi(\rho) 
\; .
\end{equation}
Here we introduced the following dimensionless quantities
\begin{equation}
        \rho = \frac{\varrho}{L}\: , \qquad
                \bar M=\frac{R^2}{L} M\: ,  \qquad
                \frac{L}{R^2} = \frac{2\pi m_q}{\sqrt{\lambda}}\: ,
\end{equation}
and identified the meson mass squared $M^2$ with the square of the momentum
four-vector $\vec k$ of the fluctuations,
\begin{equation}
        M^2=-\vec k^2
\; .
\end{equation}

The eigenfunctions $\varphi_n$ solving the Sturm-Liouville equation
\eqref{eq:eomQ0} are given in terms of the standard hypergeometric function
${}_2F_1$,
\begin{equation}
        \varphi_n(\rho) =  \frac{c_n}{(\rho^2 + 1)^{n+1}}\:{{}_2F_1\left(-(n+1); -n; 2; -\rho^2\right)} 
\; ,
\end{equation}
where $c_n$ is a normalization constant such that
\begin{equation}
\label{eq:phiOrthoNorm}
\int\limits_{0}^{\infty}\! \dd \rho\; \frac{\rho^3}{(\rho^2 + 1)^2}\,\varphi_n(\rho)\,\varphi_m(\rho) = \delta_{nm} 
\; .
\end{equation}
The lowest mode $\varphi_0$ is given by
\begin{equation}
\label{eq:phi0Solution}
        \varphi_0(\rho)=\frac{\sqrt{12}}{\rho^2+1} 
\; .
\end{equation}
The corresponding eingenvalues $M_n$ to the functions $\varphi_n$ are given by
\begin{equation}
\label{eq:massSpecQ0}
        \bar M_n = 2 \sqrt{(n+1)(n+2)} 
\; .
\end{equation}
We note that the mass of the lowest state with $n=0$ is
\begin{equation}  
\label{eq:massSpecQ0b}
   M_0  = \frac{L}{R^2}\,2\sqrt{2} = \frac{2\pi m_q}{\sqrt{\lambda}}\, 
2 \sqrt{2} 
\;  ,
\end{equation}
which will appear frequently below. For a more detailed derivation of these
results the reader is referred to \cite{Kruczenski:2003be}.

\subsubsection{Mass shift in the dilaton background}

Let us now calculate the polarizability $c_F$ which
determines the change $\delta M$ of the meson mass at a given value of the gauge
condensate $\left<\OF\right>$ with respect to the meson mass at 
$\left<\OF\right>=0$,
\begin{equation}
\label{eq:massShiftT}
        \delta M = - \frac{c_F}{N^2} \left<\OF\right> 
\; .
\end{equation}
To find $c_F$ we will determine the mass shift $\delta M$ and identify
$c_F$ with the proportionality constant in front of $\left<\OF\right>$.

We are interested in the eigenvalues of fluctuations in the case of $q \propto
\left<\OF\right> \neq 0$. The ten-dimensional background geometry dual to this
scenario is given in \eqref{eq:liuTseytlinBackground} and the equation of motion
for D$7$ brane fluctuations analog to \eqref{eq:eomQ0} was derived in
\cite{Ghoroku:2004sp} to be
\begin{equation}
\label{eq:eomQneq0}
        -\partial_\rho \rho^3 \partial_\rho \varphi(\rho) = \bar
M^2\frac{\rho^3}{(\rho^2+1)^2}\,\varphi(\rho) - 4\bar q
\frac{\rho^4}{(\rho^2 +1)(\bar q + (\rho^2 +
1)^2)}\,\partial_\rho\varphi(\rho)
\; ,
\end{equation}
with  the dimensionless $\bar q$
\begin{equation}
\label{eq:dimensionlessMandQ}
        \bar q=\frac{q}{L^4} 
\; .
\end{equation}
To obtain analytical results, we consider the case of small $\bar
q$ and linearize in this parameter. Therefore the equation of
motion to solve is
\begin{equation}
        -\partial_\rho \rho^3 \partial_\rho \varphi(\rho)
        = \bar M^2\frac{\rho^3}{(\rho^2+1)^2}\,\varphi(\rho) + \Delta(\rho) \varphi(\rho)
\; ,
\end{equation}
where the operator $\Delta(\rho)$ is given by
\begin{equation}
        \Delta(\rho) = - 4\bar q \frac{\rho^4}{(\rho^2 +1)^3}\,\partial_\rho 
\; .
\end{equation}
It is this term that describes the difference between the equation of motion at
non vanishing background perturbation to \eqref{eq:eomQ0}, which is valid for
$q=0$.

To find the solution $\varphi_0(\rho)$ corresponding to the lightest meson 
with $n=0$ we set up a perturbative expansion. Any deviation
$\delta\varphi_0$ from the solution $\varphi_0$ of the case $q=0$ may
be written as a linear combination of the functions $\varphi_n$, which are a
basis of the function space of all solutions,
\begin{align}
\label{eq:phiPerturbed}
        \phi(\rho) &=  \phi_0(\rho) + \sum_{n=0}^\infty a_n 
\phi_n(\rho) \; , & a_n &\ll 1 \; ,\\
        \bar M^2   &= \bar M_0^2 + \delta \bar M_0^2 \; , &  \delta \bar M_0^2 &\ll 1 
\; .
\end{align}
Plug this ansatz into the equation of motion \eqref{eq:eomQneq0}, make use of
\eqref{eq:eomQ0} and keep terms up to linear order in the small parameters
$a_n$, $\bar q$ and $\delta M_0^2$ to get
\begin{equation}
        \frac{\rho^3}{(\rho^2+1)^2}\,\sum_{n=0}^\infty a_n \bar M^2_n \varphi_n(\rho) =  
        \delta \bar M_0^2 \frac{\rho^3}{(\rho^2+1)^2}\,\varphi_0(\rho) + \bar M_0^2 \frac{\rho^3}{(\rho^2+1)^2} \sum_{n=0}^\infty a_n\varphi_n(\rho)
         + \Delta(\rho) \varphi_0(\rho) 
\; .
\end{equation}
We now multiply this equation by $\varphi_0(\rho)$, integrate over
$\rho \in [0,\infty]$ and make use of \eqref{eq:phiOrthoNorm} and
\eqref{eq:phi0Solution} to see that
\begin{equation}
\begin{aligned}
        \delta \bar M_0^2 &= - \int\limits_0^\infty \! \dd\rho \; \varphi_0(\rho)\, \Delta(\rho) \varphi_0(\rho) \\
                 &= -\frac{8}{5}\,\bar q
\; .
\end{aligned}
\end{equation}
From $\delta\bar M^2_0 = 2 \bar M_0 \delta\bar M_0$ 
we  obtain 
\begin{equation}
        \delta M_0 = \frac{L}{2R^2} \frac{\delta \bar M_0^2}{\bar M_0} 
        =  -\frac{8}{5\pi} \left( \frac{2\pi}{M_0} \right)^3 \,\frac{1}{N^2} \llangle \OF \rrangle
\; ,
\end{equation}
where in the last step we used \eqref{eq:massSpecQ0b} for the mass
and \eqref{eq:qandF2} and \eqref{eq:dimensionlessMandQ} to relate $\bar{q}$ and $\OF$.
By comparison with \eqref{eq:massShiftT} we  identify
the polarizability
\begin{equation}
\label{eq:alphaF}  
        c_F =  \frac{8}{5\pi} \left( \frac{2\pi}{M_0} \right)^3
\; .
\end{equation}

\subsubsection{Mass shift in the finite temperature background}

The calculation of the polarizability $c_T$ is completely analogous. We
are now looking for the proportionality constant of meson mass shifts with
respect to deviations from zero temperature,
\begin{equation}
        \label{eq:deltaMalphaT}
        \delta M = - \frac{c_T}{N^2} \llangle T^{00} \rrangle 
\; . 
\end{equation}
The background dual to the 
finite temperature field theory is the AdS black hole background given in
\eqref{eq:metricDefinitions}. Notice that the black hole radius $r_H$ is
related to the expectation value $\left<T^{00}\right>$ by \cite{Gubser:1996de}
\begin{equation}
        \left<T^{00}\right> =  \frac{3}{8} \pi^2 N^2 T^4\; , \qquad r_{H} = \pi T R^2 \,  
\; .
\end{equation}

\begin{figure}
        \includegraphics[width=.5\linewidth]{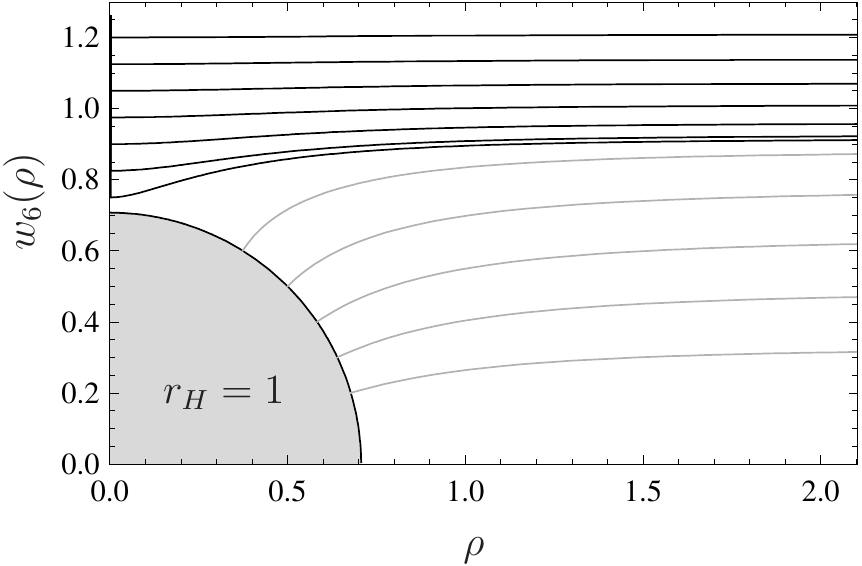}
                \caption{Some embedding functions $w_6(\rho)$ for different quark
                masses $m_q=\lim_{\rho\to\infty} w_6/(2\pi\ell_s^2)$.
                                For a detailed discussion see e.g.\ \cite{Babington:2003vm}.}
        \label{fig:BHembeddings}
\end{figure}

Again we calculate the meson mass spectrum to identify the polarizability by
comparison with \eqref{eq:deltaMalphaT}.
The embedding functions $w_5(\varrho)$ and $w_6(\varrho)$ in this background
are given by
\begin{align}
        w_5 &= 0\; ,\\
        w_6 &= w_6(\varrho)\; ,
\end{align}
where the quark mass is determined by $m_q=\lim_{\varrho\to\infty}
w_6/(2\pi\ell_s^2)$. The function $w_6(\varrho)$ has to be computed numerically
\cite{Babington:2003vm}. Some examples of such embeddings are shown in
\Fig{fig:BHembeddings}.
We introduce small fluctuations $\varphi(\rho)e^{i\vec{k}\vec{x}}$ in the $w_5$
direction,
\begin{equation}
        w_5 \to w_5(\varrho,\vec{x}) = \varphi(\varrho)e^{i\vec{k}\vec{x}}\; .
\end{equation}
The linearized equation of motion for the fluctuations $\varphi(\varrho)$ in the
limit of vanishing spatial momentum and $M^2=-\vec k^2 $ can be derived from the
DBI action \eqref{eq:DBIaction} to be
\begin{equation}
\label{eq:eomfluctLorentz}
        \begin{aligned}
                0 = & \; \hphantom{+} \partial_\varrho
                \left[ \mathcal{G} \sqrt{ \frac{1}{1 + \left( \partial_\varrho w_6  \right)^2}}\, 
                \partial_\varrho \varphi(\varrho) \right]
                 -  \sqrt{1 + \left( \partial_\varrho w_6\right)^2} \frac{\varrho^3}{2( \varrho^2 + w_6^2)^5}\, r_H^8 \,\varphi(\varrho)\\
                 & \; +
                \mathcal{G}  \sqrt{ 1 + \left( \partial_\varrho w_6 \right)^2 }\,
                \frac{4 \left( \varrho^2 +w_6^2\right)^2 + r_H^4}{ \left( (\varrho^2 +w_6^2)^2 - r_H^4\right)^2}
                \,4R^4 M^2 \varphi(\varrho) 
\; ,
        \end{aligned}
\end{equation}
where we abbreviated
\begin{equation}
        \mathcal{G} = 
                \varrho^3\left( 1-\frac{r^8_H}{16\left(\rho^2+w_6^2\right)^4} \right)
        \; .
\end{equation}

In the regime of small temperatures, we may linearize in $r_H^4$ which is the
leading order in $r_H$. Furthermore, as may be seen from
figure~\ref{fig:BHembeddings}, in the regime of a small temperature $T$ compared
to the quark mass $m_q$, or respectively small ratios of $ r_H /
\lim_{\rho\to\infty} w_6(\rho)$, the embeddings become more and more constant.
So for constant embeddings $w_6 = L$ and up to order $T^4\propto r_H^4$ the
equation of motion simplifies to
\begin{equation}
\label{eq:eomfluctLorentzLinear}
                -\partial_\rho \rho^3 \partial_\rho \varphi(\rho) = \bar M^2 \frac{\rho^3}{(\rho+1)^2}\varphi(\rho) + \Delta(\rho) \varphi(\rho) 
\; ,
\end{equation}
where we made use of the dimensionless quantities \eqref{eq:dimensionlessMandQ}
and identify
\begin{equation}
         \Delta(\rho) = \frac{3}{4}\frac{r_H^4}{L^4}\frac{\rho^3}{(\rho^2+1)^4} \bar M^2 
\; .
\end{equation}
For the lightest meson, the ansatz \eqref{eq:phiPerturbed} 
this time leads to
\begin{equation}
\begin{aligned}
        \delta \bar M_0^2 &= -\int\limits_0^\infty \! \dd\rho \, \varphi_0(\rho)\, \Delta(\rho) \varphi_0(\rho) \\
                &= -\frac{9}{40} \frac{r_H^4 \bar M_0^2}{L^4}
 \; .
\end{aligned}
\end{equation}
Reinstating units and solving for $\delta M_0$ leads to
\begin{equation}
        \delta M_0 =  -\frac{12}{5\pi} \left(\frac{2\pi}{M_0}\right)^3 \,\frac{1}{N^2} \llangle T^{00} \rrangle
\; .
\end{equation}
From this we can read off the polarizability $c_T$ as
\begin{equation}
\label{eq:alphaT}  
c_{T} = \frac{12}{5\pi}  \left(\frac{2\pi}{M_0}\right)^3
 \; .
\end{equation}

\section{Finite Temperature Correlators} 
\label{sec:adsCftCorrelators}

According to \Eq{eq:adsCftCorrelator}
we need to compute the following  correlators at 
finite temperature 
\begin{eqnarray}
G_{R}^{TT}(\omega,\q) &=& -i\int \!\dd^4x\, e^{+i\omega t - i \q\cdot \x} \,\theta(t)\llangle \left[T^{00}(\x,t) , T^{00}({\bf 0},0) \right] \rrangle 
\; , \\
G_{R}^{FF}(\omega,\q) &=& -i\int \!\dd^4x\, e^{+i\omega t - i \q\cdot \x} \,\theta(t)
\llangle\left[ \O_{F^2}(\x,t) , \O_{F^2}({\bf 0}, 0) \right] \rrangle 
\; .
\end{eqnarray} 
The calculational procedure for these two correlators is 
standard and has been discussed in \cite{Son:2002sd, Kovtun:2005ev}. $T^{00}$ correlators 
are associated with  graviton 
propagation and $\O_{F^2}$ correlators are associated with 
dilaton propagation.

On the gravity side both field  
correlators are computed in the black hole background~\eqref{eq:adsBHMetric}  
placing the dual gauge theory operator correlation functions 
at finite temperature. 
For simplicity in this section we work in the conventions and
coordinates of~\cite{Policastro:2002se}. 
We apply the method developed in~\cite{Son:2002sd,Kovtun:2005ev}
and first applied in~\cite{Policastro:2002se}, in order to find the 
two-point Minkowski correlators as   
\begin{equation}
\label{eq:retardedThermalGreen}
G^R(\omega, {\bm q})\,=\, \left.  A(u)\, f(u,-\vec k)\,
  \partial_uf(u,\vec k)\right|_{u\to 0} 
\; .    
\end{equation}
The function~$f(u,\vec k)$ relates the boundary and bulk values
of a gravity field to each other. For example the dilaton field~$\Phi$ 
is related to its value at the boundary~$\phi^{\text{bdy}}$ by    
\begin{equation}
\label{eq:rhoIs0Boundary}
\Phi(u,\vec k) = f(u,\vec k)\, \phi^{\text{bdy}}(\vec k) 
\; ,
\end{equation}
and is normalized to one at the boundary $f(0,\vec k)=1$.
For metric fluctuations, $\Phi(u,\vec k)$ is replaced by  
$h^{00}(u,\vec{k})$.
The factor~$A(u)$ can be read off from the classical supergravity 
action  
\begin{equation}
\label{eq:classicalAction}
S_{\mathrm{cl}}= \,\frac{1}{2} \int\! \mathrm d u \, \mathrm d^4x \, 
  A(u)\,(\partial_u \Phi)^2\, +\, \dots
\; .
\end{equation}
The classical gravity action for the graviton and 
dilaton is obtained from~\eqref{eq:IIBAction} as   
\begin{equation}
\label{eq:dilatonGravitonAction}
S=\frac{1}{2\kappa_{5}^2}\int\! \dd u\, \dd^{4}x\sqrt{-g_5} \left[
  (\mathcal{R}- 2\Lambda)- \frac{1}{2}(\partial \Phi)^2 
    +\dots
\right] 
\; , 
\end{equation}
where
\begin{equation}  
\frac{1}{\kappa_5^2} = \frac{R^5\Omega_5}{\kappa_{10}^2} 
  = \frac{N^2}{4\pi^2 R^3} 
\; .  
\end{equation}
So comparing to~\eqref{eq:classicalAction} we get  
\begin{equation}  
A_\Phi= -\frac{1}{2\kappa_5^2}\sqrt{-g_5} g^{uu}  
\; .    
\end{equation}   
The equation of motion derived from~\eqref{eq:dilatonGravitonAction} 
in momentum space reads  
\begin{equation} 
\label{eq:dilatonEom}  
\Phi ''-\frac{1+u^2}{u f(u)}\Phi' + 
 \frac{\wn^2-\qn^2 f(u)}{u f(u)^2}\Phi = 0
\; ,   
\end{equation} 
with the function~$f(u)=1-u^2$, the dimensionless frequency~$\wn=\omega/2\pi T$ and 
spatial momentum component~$\qn=q/2\pi T$.  
The equation of motion~\eqref{eq:dilatonEom} has to be 
solved numerically with incoming wave boundary 
condition at the black hole horizon.    
Computing the indices and expansion coefficients near the boundary as done 
in~\cite{Teaney:2006nc,Kovtun:2006pf}, we obtain the asymptotic
behavior as linear combination of two solutions    
\begin{equation} 
\label{eq:dilatonBoundary}    
\Phi(u) = (1+\dots) + \mathcal{B} (u^2+\dots) 
\; , 
\end{equation}    
where~$\mathcal{B}$ is the coefficient for the second solution 
and the coefficient for the first solution has been set to~$1$.  
At the horizon the asymptotic solution satisfying the incoming
wave boundary condition is 
\begin{equation}
\label{eq:dilatonHorizon}    
\Phi (u) = (1-u)^{-i\wn/2} (1+\dots)  
 \; . 
\end{equation}
As discussed in~\cite{Teaney:2006nc,Kovtun:2006pf} we find the 
coefficient~$\mathcal{B}$ by integrating the two boundary solutions
from~\eqref{eq:dilatonBoundary}   
forward towards the horizon and by matching the linear combination 
of the numerical solutions
$\Phi (u) = \Phi_1^{\text{num}} + \mathcal{B} \Phi_2^{\text{num}}$     
to the solution~\eqref{eq:dilatonHorizon} at the horizon. 
The imaginary part of the retarded 
correlator then is given by    
\begin{equation}  
\frac{-2 T}{\omega}\Im  G^R_{\Phi\Phi} = 
   \frac{N^2 (\pi T)^4}{4\pi^2} 
\frac{2}{\pi} \frac{\Im  \mathcal{B}}{\wn}\, .     
\end{equation}    
Solving~\eqref{eq:dilatonEom} and matching the asymptotic solutions 
as described above, we obtain 
\begin{equation}  
\label{eq:forceCorrelatorF} 
\lim_{\omega\to 0}\int\!\frac{\dd^3 \q}{(2\pi)^3}\,
  \frac{\q^2}{3}\left[\frac{-2 T}{\omega} \Im  G^R_{F^2 F^2}(\omega,\q)\right]         
  = N^2 T^9 \,67.258 
\; .   
\end{equation}    
The corresponding result for the energy-momentum tensor correlator 
is obtained in an analogous way but the analysis is significantly more 
complicated. Fortunately it has been extensively and carefully analyzed \cite{Kovtun:2005ev}. 
The final result is
\begin{equation}  
\label{eq:forceCorrelatorT} 
\lim_{\omega\to 0}\int\!\frac{\dd^3 \q}{(2\pi)^3}
  \frac{\q^2}{3}\left[\frac{-2 T}{\omega} \Im  G^R_{TT}(\omega,\q)\right]         
  = N^2 T^9\, 355.169 
\; . 
\end{equation}

\section{Summary and Discussion} \label{sec:charmDiffusion}
\label{sec:summary}

Over the duration of the lifetime of the heavy meson state the 
meson will loose momentum on average and simultaneously 
receive random kicks 
as  codified by the Langevin equations of motion
\begin{equation}  
  \frac{\dd p^i}{\dd t} = - \eta_{\scriptscriptstyle D} p^i  + \xi^i(t) 
\; ,   
\qquad  \llangle \xi^{i}(t) \xi^{j}(t') \rrangle  = \kappa \delta^{ij} \delta(t-t') 
\; .
\end{equation}  
The drag and momentum broadening rates  are related by the Einstein relation 
\begin{equation}  
\label{einstein_final}
\eta_{\scriptscriptstyle D} = \frac{\kappa}{2 T M_0}
\; ,
\end{equation}  
with $M_0$ the meson mass.
Collecting  the results for polarizabilities ~\eqref{eq:alphaF}, \eqref{eq:alphaT} 
and force correlators~\eqref{eq:forceCorrelatorF},~\eqref{eq:forceCorrelatorT}, 
and using \Eq{eq:adsCftCorrelator} we obtain our principal result
\begin{equation}  
\begin{aligned}
  \kappa &= 
\frac{T^3}{N^2} \left(\frac{2\pi T}{M_0} \right)^6 \left[
\left(\frac{8}{5\pi}\right)^2(67.258) + \left(\frac{12}{5\pi}\right)^2(355.169) \right] \\
   &= \frac{T^3}{N^2} \left(\frac{2\pi T}{M_0} \right)^6 \left[
224.726 \right]
\; .
\end{aligned}
\end{equation}  
The finite temperature mass shift in $\N=4$ SYM is
\begin{equation}  
\begin{aligned}
 \delta M_0 &= -\frac{c_{T}}{N^2} \llangle T^{00} \rrangle \\
          &= - T \, \left(\frac{2\pi T}{M_0}\right)^3\frac{9\pi}{10}
\; .
\end{aligned}
\end{equation}  
Comparing these formulas with the analogous formulas in weak coupling
large $N$ QCD given in Eqs.~\eqref{eq:kappaqcd_final} and \eqref{eq:qcd_massshift}
\begin{equation}  
   \kappa_{\mbox{\tiny pQCD}} =\, \frac{T^3}{N^2} \left(\frac{\pi T}{\Lambda_B}\right)^6
\frac{50176}{1215} \pi 
\; ,
\end{equation}  
and 
\begin{equation}  
   \delta M_{\mbox{\tiny pQCD}} =-  T\left(\frac{\pi T}{\Lambda_B} \right)^3\, \frac{14}{45} 
\; ,
\end{equation}
we see that the meson mass $M_0$ plays the role of the inverse Bohr radius
$\Lambda_B=(m_{q}/2) \alpha_s C_F$ in the strong coupling dipole effective
Lagrangian. This is as expected for relativistic bound states. Below we will
compare the values of the ratio $\kappa / (\delta M^2)$ at strong and weak
coupling.

Perhaps the theoretically most important aspect of this work is that we have
deduced a drag coefficient which is suppressed by $N^2$ in the large $N$ limit.
On the field theory side, this was achieved by calculating the mass shift of a
meson in an external background (which is finite at large $N$), and using this
information to deduce the meson couplings to the stress tensor and the operator
$\mathcal O_{F^2}$. We have restricted the calculation to the heavy dipole limit
where these are the only relevant operators. Subsequently, the fluctuations of
these operators give rise to a net force on the meson.

On the gravity side, the meson mass shift arises as a change in the normal
vibrational modes of the D$7$ brane in the presence of an external gravitational
(or dilatonic) field. Although it is not manifest in the usual black hole finite
temperature AdS/CFT setup, the gravitational field and dilatonic fields are
continually fluctuating. This is encoded by the fluctuation dissipation theorem
in the field theory which does emerge in a Kruskal formalism of the gauge
gravity duality. (The fluctuation dissipation theorem was used to relate \Eq{ff}
and \Eq{ff2}.) Since these fluctuating gravitational and dilatonic fields shift
the spectrum of the D7 brane excitations, gradients in these fields give rise to
a net force on a mesonic normal modes of the D7 brane. This discussion suggests
a better understanding of how gravitational and dilatonic fields fluctuate in
bulk, would give a straightforward procedure to calculate the drag of a
finite mass meson. Specifically, fluctuations in the bulk would force motion of
meson wave functions which extend into the fifth dimension. We hope to pursue
this reasoning in the future.

From a phenomenological perspective the current calculation was limited to very
heavy mesons (which survive above $T_c$) where dipole interactions between the
meson and the medium are dominant. It is certainly unclear if this is the
relevant interaction mechanism above $T_c$ even for bottomonium. Furthermore, the
dipole coupling between a heavy meson and the medium is dominated by short
distance physics which is not well modeled by AdS/CFT.

\begin{figure}
	\begin{center}
	\includegraphics[width=2.3in]{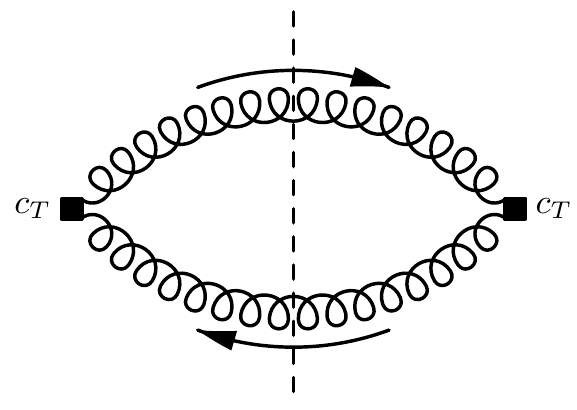}
	\caption{
		Feynman graph leading to scattering of a heavy meson in perturbation
		theory. This graph represents  the gauge contribution to correlators
		given in \Eq{eq:strong}.  At strong coupling the effect of
		additional scatterings is to reduce the integrated value of this 
		correlator by almost a factor of five in the appropriate kinematic 
		regime.}
	\label{graph2}
	\end{center}
\end{figure}

However, after the gluons scatter off the heavy quark, they propagate out into
plasma which modifies the free propagation as indicated by the Feynman graph in
\Fig{graph2}. To factorize this long distance dynamics from the short distance
meson dynamics we form the ratio
\begin{equation}
\label{eq:strong}
\begin{aligned}
\left[ \frac{\kappa}{(\delta M)^2} \right]_{\lambda \rightarrow \infty} &\simeq \frac{-1}{(\llangle T^{00} \rrangle)^2 }  \int_{-\infty}^{\infty} \dd t   \, 
\left. \nabla^2_{\bf y}  \llangle T^{00}({\bf y},t)T^{00}(\x,0) \rrangle \right|_{{\bf y}=\x} \, ,  \\
 &\simeq  \frac{\pi T}{N^2}\,  8.25 
\; ,
\end{aligned}
\end{equation}
which is independent of the short distance coefficients $c_T$ and $c_F$ provided
the numerically small dilatonic contribution is neglected\footnote{With the
dilaton contribution the coefficient is 8.95. }. It is then reasonable to use
AdS/CFT to estimate to what degree strong coupling physics modifies this ratio
in QCD. In the free finite temperature $\N=4$ theory, the result is (see
\App{perturb})
\begin{equation}   \label{eq:weak2}
\left[ \frac{\kappa}{(\delta M)^2} \right]_{
\lambda\rightarrow 0} \simeq  \frac{\pi T}{N^2} \, 37.0
\; ,
\end{equation}
Thus comparing the strong coupling result \eqref{eq:strong} with the
weak-coupling result \eqref{eq:weak2}, we conclude that strong coupling effects
actually {\it reduce} the scattering rate relative to the mass shift. Roughly
speaking, the same strong coupling physics that is responsible for the reduction
of pressure (by a factor of $3/4$) relative to the Stefan-Boltzmann prediction
is at work here. However the effect is more pronounced since the correlator in
\Eq{eq:strong} is dominated by larger values of spatial momentum $\q$.

Given this AdS/CFT result we expect the perturbative
estimates for the rate of momentum broadening  to be \emph{reduced}
by some factor which could be as large as a factor of five. We
will not speculate on this factor here but simply 
write the perturbative momentum diffusion rate as
\begin{equation} 
\label{final_analytic}
  \kappa_{\text{\tiny QCD} }  = T (\delta M)^2 \,  \frac{1280\pi}{3 N^2} 
\; ,
\end{equation}  
with a gravitionally biased opinion  that the coefficient  is too large.
To obtain a numerical estimate for 
the $\Upsilon(1s)$ state, we take $T=340\,{\rm  MeV} \simeq 2 T_c$, \,
$M_0 = 9.46\,{\rm GeV}$,
and estimate the mass shift as  $\delta M \simeq -10 \, \mbox{MeV}$ 
based on potential model calculations
 which fit lattice data \cite{Mocsy:2007yj,Mocsy:2007jz}. Substituting   
into \Eq{final_analytic} and \Eq{einstein_final} for the relaxation time $\tau_R \equiv \eta_{\scriptscriptstyle D}^{-1}$
we find
\begin{equation}  
\label{final1}
   \kappa_{\text{\tiny QCD} } = 0.025\, 
\frac{{\rm GeV}^2}{\rm fm} \, 
\left(\frac{T}{340\,{\rm MeV} } \right) \left( \frac{\delta M_0}{
10\, {\rm MeV} }\right)^2
\; ,
\end{equation} 
\begin{equation}  
\label{final2}
   \tau_{R} \equiv \frac{1}{\eta_{\scriptscriptstyle D}} = 250\, 
{\rm fm}
\left(\frac{M_0}{9.46\,{\rm GeV}} \right) \left( \frac{10\,{\rm MeV}}
{\delta M_0} \right)^2 
\; . 
\end{equation}  
The introduction discusses some of the significant uncertainties
associated with the current formalism and the extraction of mass shift 
adopted here.
Nevertheless, the AdS/CFT prediction is that strong coupling effects
will actually increase  this perturbative
estimate of the  relaxation time by up to a factor of five.


\begin{acknowledgments}
\label{sec:ack}
We especially thank P.~Petreczky for extensive discussion, 
for providing numerical values leading to \Eq{final1} and \Eq{final2}.
We thank M.~Ammon, T.~Faulkner, H.~Liu, O.~Philipsen and A.~Tseytlin
for discussions related to this work, as well as G.~Korchemsky and E.~Kiritsis
for useful comments.
Derek Teaney is supported  in part by an OJI grant from 
the U.S. Department of Energy and the Sloan Foundation. 
Johanna Erdmenger, Matthias Kaminski and Felix Rust are supported in part by
\emph{The Cluster of Excellence for Fundamental Physics -- Origin and Structure
  of the Universe.}

\end{acknowledgments}

\begin{appendix}

\section{Diffusion Rate in Perturbation Theory}
\label{perturb}

The purpose of this appendix is to compute $\kappa/(\delta M)^2$ in free finite
temperature field theory in $\N=4$ Super Yang Mills theory. This will permit a
comparison to the strongly interacting results.

\subsection{QCD Computation}

\begin{figure}
\begin{center}
\includegraphics[width=2.7in]{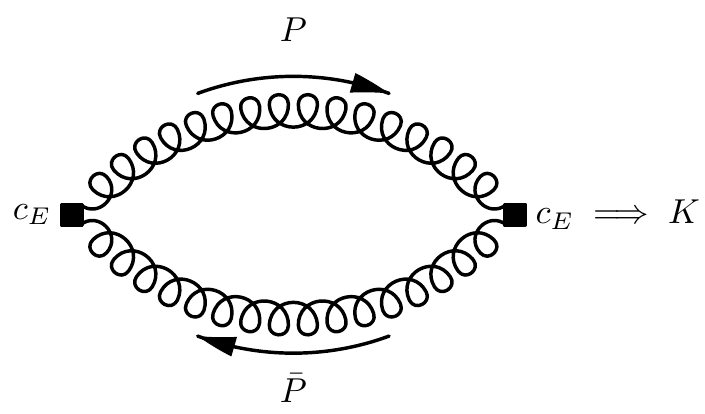}
\end{center}
\caption{
\label{graph3}
Euclidean graph corresponding to the \Eq{e2e2formula} and \Eq{e2e2graph}.
}
\end{figure}

In the interest of pedagogy we will indicate in detail how the QCD computation
is performed. We will work in the limit where only the coupling to the electric
field is included $i.e.$, $c_{B}=0$. $\kappa$ is given by \Eq{e2e2formula}. We
will use the Matsubara formalism though the real time formalism is not more
difficult in this case. We will work in the Coulomb gauge where the propogators
are
\bg
 \int_0^{\beta}\!\dd\tau \int\! \dd^3\x\, e^{-iK\cdot X} \llangle  A_{0}(X) A_{0}(0) \rrangle 
&=& \frac{1}{k^2} 
\; , \\
 \int_0^{\beta}\!\dd\tau \int\! \dd^3\x\, e^{-iK\cdot X} \llangle  A_{i}(X) A_{j}(0) \rrangle  &=& \frac{\hat k_i\hat k_j -\delta_{ij}}{K^2}
\; .
\nd
Here we follow standard thermal field theory notation $K^{\mu}=(\omega_n,\k)$
and $X^{\mu} = (\tau,\x)$ with $k=\left|\k\right|$. $\omega_n=2\pi n T$ labels
the Matsubara and $K\cdot X = \omega_n \tau + \k\cdot\x$. The Euclidean metric
is $g^{\mu\nu}_{E}={\rm diag}(+,+,+,+)$ and further explanation of Euclidean
conventions is given in \Ref{Petreczky:2005nh} . Then the Euclidean correlator
corresponding to \Eq{e2e2formula} and \Fig{graph3} for a single color index
\begin{equation} 
\label{e2e2graph}
 G^{\E^2\E^2}_{E}(K) = \frac{1}{2} T\sum_{P^0}\! \int\! \frac{\dd^3\p}{(2\pi)^3}  
 \, P^0 \bar{P}^0\, \frac{\hat p_i\hat p_j  -\delta_{ij}}{P^2}\, P^0\bar{P}^0\,\frac{\hat\pbar_j \hat\pbar_i - \delta_{ji}}{\bar P^2} 
+ \mbox{ Coulomb graphs} 
\end{equation}  
here $\bar{P} = P - K$ and we do not write the graphs involving 
coulomb lines since these do not contribute to the imaginary part.
Performing the Matsubara sum, 
analytically continuing $-iK^{0}\rightarrow \omega + i\epsilon$, 
taking the 
imaginary part, and finally working
in the limit that $\omega\rightarrow0$ yields the following 
result for the imaginary part of the retarded correlator 
\begin{equation}  
\lim_{\omega\rightarrow 0} 
\frac{-2T}{\omega} 
\Im G_{R}^{\E^2\E^2} (\omega,\k) =
\int 
\! \frac{\dd^3p}{(2\pi)^3 2E_\p} 
\frac{\dd^3\pbar}{(2\pi)^3 2E_{\bf \pbar}} \, 
n_\p(1 + n_\p)
 2\pi \delta(E_\p - E_{\bf \pbar}) 
(2\pi)^3\delta^3(\p - {\bf \pbar} -\k) \left|\M\right|^2
\; ,
\end{equation}  
with 
\begin{equation}  
\left|\M\right|^2 = (E_\p E_{{\bf \pbar}})^2 \, \left(1 + \cos^2(\theta_{\p\bf\pbar} \right)
\; .
\end{equation}
The details of the preceding steps can be streamlined and are 
found in many 
places; see \Ref{Petreczky:2005nh}  and the text book \cite{Lebellac} for simple explanations.
 Integrating over
the retarded correlator as required by \Eq{e2e2formula}  and
multiplying by $N^2$ to account for the number of gluons yields the following 
result 
\begin{equation}  
\kappa = \frac{c_{E}^2}{N^2} \frac{64\pi^5}{135} T^9
\; .
\end{equation}  
This result is the expected kinetic theory result for the 
rate of momentum diffusion of a heavy meson scattering  via 
dipole scattering.

\subsection{$\N=4$ Computation}

\label{free_section}
The free $\N=4$ Lagrangian is written as follows:
\begin{equation}
\L = 2 \tr \left\{
-\frac{1}{4}F^2 + \frac{1}{2} \bar{\lambda}_a (-i
\bar{\sigma} \cdot \partial )  \lambda^a - \frac{1}{2} \partial_{\mu} X_i \partial^{\mu}X_i  \right\} 
\; ,
\end{equation}
where ``$a$'' is a SU(4) index and ``$i$'' is a SO(6) index. 
Under flavor rotation,  $\lambda^a$ transforms
in the fundamental representation of SU(4) and 
$X^i$ transforms as the fundamental representation of 
SO(6). SU(4) and SO(6) are locally isomorphic. 
SU(4) matrices are parameterized as $e^{i\beta_A (T^4)_A}$ 
with trace normalization $\tr[(T^4)_A (T^4)_B] = C_4 \delta_{AB}$ and $C_4 =1/2$.
Similarly,  SO(6) matrices are written as $e^{i\beta_A (T^6)_A}$,
with trace normalization $C_6=1$ \cite{Peskin}. The 
normalization convention adopted here has been fixed 
so that the AdS/CFT correspondence holds at the level of 
non-renormalized two 
point functions at zero temperature \cite{Freedman:1998tz}.

The full stress tensor is written
\begin{equation}  
 T^{\mu\nu} = \left(T^{\mu\nu}\right)_{\rm gauge}
                + \left(T^{\mu\nu}\right)_{\rm fermion}
                 + \left(T^{\mu\nu}\right)_{\rm scalar} 
\; ,
\end{equation}  
with
\bg
  \left(T^{\mu\nu}\right)_{\rm gauge}  &=&  2\, \tr \left\{ F^{\mu}_{\kappa} F^{\nu \kappa} 
  + g^{\mu\nu}\left( -\frac{1}{4}F^2 \right) \right\} \; ,
\label{freerhoa} \\
  \left(T^{\mu\nu}\right)_{\rm fermion} &=& 
2\,\tr\left\{ 
\frac{i}{8} 
\bar{\lambda} \left( \bar{\sigma}^{\mu} \overleftrightarrow{\partial^\nu}  + \bar{\sigma}^{\nu} \overleftrightarrow{\partial^\mu} \right) \lambda + 
g^{\mu\nu} \left(\frac{1}{2}
\bar{\lambda}^{a}(-i\bar{\sigma}\cdot \partial) \lambda_a \right) \right\} 
\; , \\
  \left(T^{\mu\nu}\right)_{\rm scalar} &=& 
2\,\tr\left\{ 
          \partial^{\mu}X_{i} \partial^{\nu} X_{i} +
   g^{\mu\nu} \left(-\frac{1}{2} \partial_{\alpha} X_i \partial^{\alpha} 
X_i \right) 
\right\}
\; .
\label{freerhob}
\nd
Here $\overleftrightarrow{\partial} \equiv \overrightarrow{\partial} - \overleftarrow{\partial}$.

There are three graphs for the gauge, fermion, and scalar  loops
which make a contribution to 
the imaginary part of the retarded correlator. 
The full diffusion rate is  
\begin{equation} 
 \kappa = (\kappa)_{A} + (\kappa)_\lambda + (\kappa)_X
\; ,
\end{equation}  
where $(\kappa)_{A}$ is due to gauge bosons, 
$(\kappa)_\lambda$ is due to fermions and $(\kappa)_X$ 
is due to  scalars.   In each case the retarded correlator
can be written in the form of a phase space integral 
times a matrix element squared 
The matrix elements are
\bg
\left|\M\right|^2_{A} &=& 
\left[N^2\right] \, E_\p^4 (1 + \cos(\theta_{\p\bf\pbar}))^2 (1 + \cos^2(\theta_{\p\bf\pbar}) ) 
\; , \\
\left|\M\right|^2_{\lambda} &=& 
[4 N^2] \, 4E_{p}^4 \left(1 + \cos(\theta_{\p\bf\pbar}) \right) 
\; , \\
\left|\M\right|^2_{X}  &=& 
[6 N^2 ]\, E_p^4 (1 + \cos\theta_{\p\bf\pbar})^2 
\; .
\nd
Then integrating over the phase-space we obtain the three contributions
to $\kappa$ 
\begin{eqnarray}
\left(\frac{N^4}{c_T^2} \kappa \right)_{A} &=& 
[N^2] \frac{64\pi^5}{225} T^9 
\; , \\
\left(\frac{N^4}{c_T^2} \kappa \right)_{\lambda} &=& 
[4 N^2] \frac{254\pi^5}{135} T^9 
\; , \\
\left(\frac{N^4}{c_T^2} \kappa \right)_{X} &=& 
[6 N^2]\frac{32 \pi^5}{135} T^9
\; . 
\end{eqnarray}
The final result for the momentum diffusion rate when only 
the stress tensor coupling is included is 
\bg
\left(\frac{N^4}{c_{T}^2}\kappa\right) &=& N^2\frac{6232 \pi^5}{675} T^9
\; .
\nd

Similarly the mass shift for the meson in the finite temperature 
background is  
\bg
\delta M = (\delta M)_{A} + (\delta M)_{\lambda} + (\delta M)_{X}
\; .
\nd
The different components of the mass shift are
\bg
(\delta M)_{A} &=& \frac{c_{T}}{N^2}  [2 N^2]  \frac{\pi^2 T^4}{30} 
\; , \\
(\delta M)_{\lambda} &=& \frac{c_T}{N^2} [8N^2\,\frac{7}{8}] \frac{\pi^2 T^4}{30} 
\; , \\
(\delta M)_{X}  &=& \frac{c_{T}}{N^2} [6N^2] \frac{\pi^2 T^4}{30}
\; .
\nd
In each case the $\pi^2 T^4/30$ is the energy density of a massless
single component bose gas. The factor in square brackets counts
the number of degrees of freedom (including spin) and a 
factor of $7/8$ to account for the differences between bose and fermi 
distributions. Putting these pieces together we find the 
total mass shift due to coupling to the tensor
\bg
 \delta M = c_{T} \frac{\pi^2 T^4}{2}
\; .
\nd 
Now we finally evaluate the ratio in the free theory
\begin{equation} 
\begin{aligned}
  \frac{\kappa}{(\delta M)^2} &= \,\frac{24928}{675} \frac{\pi T}{N^2} \\
                              &\simeq 37.0 \frac{\pi T}{N^2}
 \; .
\end{aligned}
\end{equation} 
This is the weak coupling expectation for this ratio provided the dominant coupling of the medium to the dipole is through the stress tensor operator.  It is useful to compare this expectation to the strong coupling results as is done in the body of the text.

\end{appendix}  

\bibliographystyle{./felice_utcaps}
\bibliography{bibliography}

\providecommand{\href}[2]{#2}\begingroup\raggedright\begin{thebibliography}{10}

\bibitem{Adare:2006nq}
{\bf PHENIX} Collaboration, A.~Adare {\em et al.}, \emph{ {Energy Loss and Flow
  of Heavy Quarks in Au+Au Collisions at $sqrt{s_{NN}}$ = 200 GeV}}, Phys. Rev.
  Lett. {\bf 98} (2007) 172301,
\href{http://www.slac.stanford.edu/spires/find/hep/www?texkey=Adare:2006nq}{nu%
cl-ex/0611018}.

\bibitem{Bielcik:2005wu}
{\bf STAR} Collaboration, J.~Bielcik, \emph{ {Centrality dependence of heavy
  flavor production from single electron measurement in s(NN)**(1/2) = 200-GeV
  Au + Au collisions}}, Nucl. Phys. {\bf A774} (2006) 697--700,
\href{http://www.slac.stanford.edu/spires/find/hep/www?texkey=Bielcik:2005wu}{%
nucl-ex/0511005}.

\bibitem{Adare:2008sh}
{\bf PHENIX} Collaboration, A.~Adare {\em et al.}, \emph{ {J/psi Production in
  $sqrt{s_{NN}}$= 200 GeV Cu+Cu Collisions}},
\href{http://www.slac.stanford.edu/spires/find/hep/www?texkey=Adare:2008sh}{08%
01.0220}.

\bibitem{Adare:2006ns}
{\bf PHENIX} Collaboration, A.~Adare {\em et al.}, \emph{ {J/psi production vs
  centrality, transverse momentum, and rapidity in Au + Au collisions at
  s(NN)**(1/2) = 200- GeV}}, Phys. Rev. Lett. {\bf 98} (2007) 232301,
\href{http://www.slac.stanford.edu/spires/find/hep/www?texkey=Adare:2006ns}{nu%
cl-ex/0611020}.

\bibitem{Adler:2005ph}
{\bf PHENIX} Collaboration, S.~S. Adler {\em et al.}, \emph{ {J/psi production
  and nuclear effects for d + Au and p + p collisions at s(NN)**(1/2) =
  200-GeV}}, Phys. Rev. Lett. {\bf 96} (2006) 012304,
\href{http://www.slac.stanford.edu/spires/find/hep/www?texkey=Adler:2005ph}{nu%
cl-ex/0507032}.

\bibitem{Arnaldi:2006ee}
{\bf NA60} Collaboration, R.~Arnaldi {\em et al.}, \emph{ {Anomalous J/psi
  suppression in In-In collisions at 158- GeV/nucleon}}, Nucl. Phys. {\bf A774}
  (2006)
711--714.

\bibitem{Alessandro:2004ap}
{\bf NA50} Collaboration, B.~Alessandro {\em et al.}, \emph{ {A new measurement
  of J/psi suppression in Pb - Pb collisions at 158-GeV per nucleon}}, Eur.
  Phys. J. {\bf C39} (2005) 335--345,
\href{http://www.slac.stanford.edu/spires/find/hep/www?texkey=Alessandro:2004a%
p}{hep-ex/0412036}.

\bibitem{Maldacena:1997re}
J.~M. Maldacena, \emph{ {The large N limit of superconformal field theories and
  supergravity}}, Adv. Theor. Math. Phys. {\bf 2} (1998) 231--252,
\href{http://www.slac.stanford.edu/spires/find/hep/www?texkey=Maldacena:1997re%
}{hep-th/9711200}.

\bibitem{Witten:1998qj}
E.~Witten, \emph{ {Anti-de Sitter space and holography}}, Adv. Theor. Math.
  Phys. {\bf 2} (1998) 253--291,
\href{http://www.slac.stanford.edu/spires/find/hep/www?texkey=Witten:1998qj}{h%
ep-th/9802150}.

\bibitem{Gubser:1998bc}
S.~S. Gubser, I.~R. Klebanov, and A.~M. Polyakov, \emph{ {Gauge theory
  correlators from non-critical string theory}}, Phys. Lett. {\bf B428} (1998)
  105--114,
\href{http://www.slac.stanford.edu/spires/find/hep/www?texkey=Gubser:1998bc}{h%
ep-th/9802109}.

\bibitem{Aharony:1999ti}
O.~Aharony, S.~S. Gubser, J.~M. Maldacena, H.~Ooguri, and Y.~Oz, \emph{ {Large
  N field theories, string theory and gravity}}, Phys. Rept. {\bf 323} (2000)
  183--386,
\href{http://www.slac.stanford.edu/spires/find/hep/www?texkey=Aharony:1999ti}{%
hep-th/9905111}.

\bibitem{Herzog:2006gh}
C.~P. Herzog, A.~Karch, P.~Kovtun, C.~Kozcaz, and L.~G. Yaffe, \emph{ {Energy
  loss of a heavy quark moving through N = 4 supersymmetric Yang-Mills
  plasma}}, JHEP {\bf 07} (2006) 013,
\href{http://www.slac.stanford.edu/spires/find/hep/www?texkey=Herzog:2006gh}{h%
ep-th/0605158}.

\bibitem{CasalderreySolana:2006rq}
J.~Casalderrey-Solana and D.~Teaney, \emph{ {Heavy quark diffusion in strongly
  coupled N = 4 Yang Mills}}, Phys. Rev. {\bf D74} (2006) 085012,
\href{http://www.slac.stanford.edu/spires/find/hep/www?texkey=CasalderreySolan%
a:2006rq}{hep-ph/0605199}.

\bibitem{Gubser:2006bz}
S.~S. Gubser, \emph{ {Drag force in AdS/CFT}}, Phys. Rev. {\bf D74} (2006)
  126005,
\href{http://www.slac.stanford.edu/spires/find/hep/www?texkey=Gubser:2006bz}{h%
ep-th/0605182}.

\bibitem{Liu:2006nn}
H.~Liu, K.~Rajagopal, and U.~A. Wiedemann, \emph{ {An AdS/CFT calculation of
  screening in a hot wind}}, Phys. Rev. Lett. {\bf 98} (2007) 182301,
\href{http://www.slac.stanford.edu/spires/find/hep/www?texkey=Liu:2006nn}{hep-%
ph/0607062}.

\bibitem{Peeters:2006iu}
K.~Peeters, J.~Sonnenschein, and M.~Zamaklar, \emph{ {Holographic melting and
  related properties of mesons in a quark gluon plasma}}, Phys. Rev. {\bf D74}
  (2006) 106008,
\href{http://www.slac.stanford.edu/spires/find/hep/www?texkey=Peeters:2006iu}{%
hep-th/0606195}.

\bibitem{Ejaz:2007hg}
Q.~J. Ejaz, T.~Faulkner, H.~Liu, K.~Rajagopal, and U.~A. Wiedemann, \emph{ {A
  limiting velocity for quarkonium propagation in a strongly coupled plasma via
  AdS/CFT}}, JHEP {\bf 04} (2008) 089,
\href{http://www.slac.stanford.edu/spires/find/hep/www?texkey=Ejaz:2007hg}{071%
2.0590}.

\bibitem{Mateos:2007vn}
D.~Mateos, R.~C. Myers, and R.~M. Thomson, \emph{ {Thermodynamics of the
  brane}}, JHEP {\bf 05} (2007) 067,
\href{http://www.slac.stanford.edu/spires/find/hep/www?texkey=Mateos:2007vn}{h%
ep-th/0701132}.

\bibitem{Myers:2008cj}
R.~C. Myers and A.~Sinha, \emph{ {The fast life of holographic mesons}},
\href{http://www.slac.stanford.edu/spires/find/hep/www?texkey=Myers:2008cj}{08%
04.2168}.

\bibitem{Erdmenger:2007ja}
J.~Erdmenger, M.~Kaminski, and F.~Rust, \emph{ {Holographic vector mesons from
  spectral functions at finite baryon or isospin density}}, Phys. Rev. {\bf
  D77} (2008) 046005,
\href{http://www.slac.stanford.edu/spires/find/hep/www?texkey=Erdmenger:2007ja%
}{0710.0334}.

\bibitem{Hoyos:2006gb}
C.~Hoyos-Badajoz, K.~Landsteiner, and S.~Montero, \emph{ {Holographic Meson
  Melting}}, JHEP {\bf 04} (2007) 031,
\href{http://www.slac.stanford.edu/spires/find/hep/www?texkey=Hoyos:2006gb}{he%
p-th/0612169}.

\bibitem{Myers:2007we}
R.~C. Myers, A.~O. Starinets, and R.~M. Thomson, \emph{ {Holographic spectral
  functions and diffusion constants for fundamental matter}}, JHEP {\bf 11}
  (2007) 091,
\href{http://www.slac.stanford.edu/spires/find/hep/www?texkey=Myers:2007we}{07%
06.0162}.

\bibitem{Erdmenger:2008yj}
J.~Erdmenger, M.~Kaminski, P.~Kerner, and F.~Rust, \emph{ {Finite baryon and
  isospin chemical potential in AdS/CFT with flavor}},
\href{http://www.slac.stanford.edu/spires/find/hep/www?texkey=Erdmenger:2008yj%
}{0807.2663}.

\bibitem{Faulkner:2008qk}
T.~Faulkner and H.~Liu, \emph{ {Meson widths from string worldsheet
  instantons}},
\href{http://www.slac.stanford.edu/spires/find/hep/www?texkey=Faulkner:2008qk}%
{0807.0063}.

\bibitem{Yao:2006px}
{\bf Particle Data Group} Collaboration, W.~M. Yao {\em et al.}, \emph{ {Review
  of particle physics}}, J. Phys. {\bf G33} (2006)
1--1232.

\bibitem{Aoki:2006br}
Y.~Aoki, Z.~Fodor, S.~D. Katz, and K.~K. Szabo, \emph{ {The QCD transition
  temperature: Results with physical masses in the continuum limit}}, Phys.
  Lett. {\bf B643} (2006) 46--54,
\href{http://www.slac.stanford.edu/spires/find/hep/www?texkey=Aoki:2006br}{hep%
-lat/0609068}.

\bibitem{Cheng:2006qk}
M.~Cheng {\em et al.}, \emph{ {The transition temperature in QCD}}, Phys. Rev.
  {\bf D74} (2006) 054507,
\href{http://www.slac.stanford.edu/spires/find/hep/www?texkey=Cheng:2006qk}{he%
p-lat/0608013}.

\bibitem{Laine:2006ns}
M.~Laine, O.~Philipsen, P.~Romatschke, and M.~Tassler, \emph{ {Real-time static
  potential in hot QCD}}, JHEP {\bf 03} (2007) 054,
\href{http://www.slac.stanford.edu/spires/find/hep/www?texkey=Laine:2006ns}{he%
p-ph/0611300}.

\bibitem{Laine:2007gj}
M.~Laine, \emph{ {A resummed perturbative estimate for the quarkonium spectral
  function in hot QCD}}, JHEP {\bf 05} (2007) 028,
\href{http://www.slac.stanford.edu/spires/find/hep/www?texkey=Laine:2007gj}{07%
04.1720}.

\bibitem{Laine:2007qy}
M.~Laine, O.~Philipsen, and M.~Tassler, \emph{ {Thermal imaginary part of a
  real-time static potential from classical lattice gauge theory simulations}},
  JHEP {\bf 09} (2007) 066,
\href{http://www.slac.stanford.edu/spires/find/hep/www?texkey=Laine:2007qy}{07%
07.2458}.

\bibitem{Burnier:2007qm}
Y.~Burnier, M.~Laine, and M.~Vepsalainen, \emph{ {Heavy quarkonium in any
  channel in resummed hot QCD}}, JHEP {\bf 01} (2008) 043,
\href{http://www.slac.stanford.edu/spires/find/hep/www?texkey=Burnier:2007qm}{%
0711.1743}.

\bibitem{Brambilla:2008cx}
N.~Brambilla, J.~Ghiglieri, A.~Vairo, and P.~Petreczky, \emph{ {Static
  quark-antiquark pairs at finite temperature}},
\href{http://www.slac.stanford.edu/spires/find/hep/www?texkey=Brambilla:2008cx%
}{0804.0993}.

\bibitem{Umeda:2002vr}
T.~Umeda, K.~Nomura, and H.~Matsufuru, \emph{ {Charmonium at finite temperature
  in quenched lattice QCD}}, Eur. Phys. J. {\bf C39S1} (2005) 9--26,
\href{http://www.slac.stanford.edu/spires/find/hep/www?texkey=Umeda:2002vr}{he%
p-lat/0211003}.

\bibitem{Asakawa:2003re}
M.~Asakawa and T.~Hatsuda, \emph{ {J/psi and eta/c in the deconfined plasma
  from lattice QCD}}, Phys. Rev. Lett. {\bf 92} (2004) 012001,
\href{http://www.slac.stanford.edu/spires/find/hep/www?texkey=Asakawa:2003re}{%
hep-lat/0308034}.

\bibitem{Datta:2003ww}
S.~Datta, F.~Karsch, P.~Petreczky, and I.~Wetzorke, \emph{ {Behavior of
  charmonium systems after deconfinement}}, Phys. Rev. {\bf D69} (2004) 094507,
\href{http://www.slac.stanford.edu/spires/find/hep/www?texkey=Datta:2003ww}{he%
p-lat/0312037}.

\bibitem{Iida:2006mv}
H.~Iida, T.~Doi, N.~Ishii, H.~Suganuma, and K.~Tsumura, \emph{ {Charmonium
  properties in deconfinement phase in anisotropic lattice QCD}}, Phys. Rev.
  {\bf D74} (2006) 074502,
\href{http://www.slac.stanford.edu/spires/find/hep/www?texkey=Iida:2006mv}{hep%
-lat/0602008}.

\bibitem{Jakovac:2006sf}
A.~Jakovac, P.~Petreczky, K.~Petrov, and A.~Velytsky, \emph{ {Quarkonium
  correlators and spectral functions at zero and finite temperature}}, Phys.
  Rev. {\bf D75} (2007) 014506,
\href{http://www.slac.stanford.edu/spires/find/hep/www?texkey=Jakovac:2006sf}{%
hep-lat/0611017}.

\bibitem{Aarts:2007pk}
G.~Aarts, C.~Allton, M.~B. Oktay, M.~Peardon, and J.-I. Skullerud, \emph{
  {Charmonium at high temperature in two-flavor QCD}}, Phys. Rev. {\bf D76}
  (2007) 094513,
\href{http://www.slac.stanford.edu/spires/find/hep/www?texkey=Aarts:2007pk}{07%
05.2198}.

\bibitem{Mocsy:2007yj}
A.~Mocsy and P.~Petreczky, \emph{ {Can quarkonia survive deconfinement ?}},
  Phys. Rev. {\bf D77} (2008) 014501,
\href{http://www.slac.stanford.edu/spires/find/hep/www?texkey=Mocsy:2007yj}{07%
05.2559}.

\bibitem{Mocsy:2007jz}
A.~Mocsy and P.~Petreczky, \emph{ {Color Screening Melts Quarkonium}}, Phys.
  Rev. Lett. {\bf 99} (2007) 211602,
\href{http://www.slac.stanford.edu/spires/find/hep/www?texkey=Mocsy:2007jz}{07%
06.2183}.

\bibitem{Karch:2002sh}
A.~Karch and E.~Katz, \emph{ {Adding flavor to AdS/CFT}}, JHEP {\bf 06} (2002)
  043,
\href{http://www.slac.stanford.edu/spires/find/hep/www?texkey=Karch:2002sh}{he%
p-th/0205236}.

\bibitem{Kruczenski:2003be}
M.~Kruczenski, D.~Mateos, R.~C. Myers, and D.~J. Winters, \emph{ {Meson
  spectroscopy in AdS/CFT with flavour}}, JHEP {\bf 07} (2003) 049,
\href{http://www.slac.stanford.edu/spires/find/hep/www?texkey=Kruczenski:2003b%
e}{hep-th/0304032}.

\bibitem{Babington:2003vm}
J.~Babington, J.~Erdmenger, N.~J. Evans, Z.~Guralnik, and I.~Kirsch, \emph{
  {Chiral symmetry breaking and pions in non-supersymmetric gauge / gravity
  duals}}, Phys. Rev. {\bf D69} (2004) 066007,
\href{http://www.slac.stanford.edu/spires/find/hep/www?texkey=Babington:2003vm%
}{hep-th/0306018}.

\bibitem{Constable:1999ch}
N.~R. Constable and R.~C. Myers, \emph{ {Exotic scalar states in the AdS/CFT
  correspondence}}, JHEP {\bf 11} (1999) 020,
\href{http://www.slac.stanford.edu/spires/find/hep/www?texkey=Constable:1999ch%
}{hep-th/9905081}.

\bibitem{Kirsch:2004km}
I.~Kirsch, \emph{ {Generalizations of the AdS/CFT correspondence}}, Fortsch.
  Phys. {\bf 52} (2004) 727--826,
\href{http://www.slac.stanford.edu/spires/find/hep/www?texkey=Kirsch:2004km}{h%
ep-th/0406274}.

\bibitem{Kruczenski:2003uq}
M.~Kruczenski, D.~Mateos, R.~C. Myers, and D.~J. Winters, \emph{ {Towards a
  holographic dual of large-N(c) QCD}}, JHEP {\bf 05} (2004) 041,
\href{http://www.slac.stanford.edu/spires/find/hep/www?texkey=Kruczenski:2003u%
q}{hep-th/0311270}.

\bibitem{Mateos:2006nu}
D.~Mateos, R.~C. Myers, and R.~M. Thomson, \emph{ {Holographic phase
  transitions with fundamental matter}}, Phys. Rev. Lett. {\bf 97} (2006)
  091601,
\href{http://www.slac.stanford.edu/spires/find/hep/www?texkey=Mateos:2006nu}{h%
ep-th/0605046}.

\bibitem{Erdmenger:2007cm}
J.~Erdmenger, N.~Evans, I.~Kirsch, and E.~Threlfall, \emph{ {Mesons in
  Gauge/Gravity Duals - A Review}}, Eur. Phys. J. {\bf A35} (2008) 81--133,
\href{http://www.slac.stanford.edu/spires/find/hep/www?texkey=Erdmenger:2007cm%
}{0711.4467}.

\bibitem{Liu:1999fc}
H.~Liu and A.~A. Tseytlin, \emph{ {D3-brane D-instanton configuration and N = 4
  super YM theory in constant self-dual background}}, Nucl. Phys. {\bf B553}
  (1999) 231--249,
\href{http://www.slac.stanford.edu/spires/find/hep/www?texkey=Liu:1999fc}{hep-%
th/9903091}.

\bibitem{Luke:1992tm}
M.~E. Luke, A.~V. Manohar, and M.~J. Savage, \emph{ {A QCD Calculation of the
  interaction of quarkonium with nuclei}}, Phys. Lett. {\bf B288} (1992)
  355--359,
\href{http://www.slac.stanford.edu/spires/find/hep/www?texkey=Luke:1992tm}{hep%
-ph/9204219}.

\bibitem{Peskin:1979va}
M.~E. Peskin, \emph{ {Short Distance Analysis for Heavy Quark Systems. 1.
  Diagrammatics}}, Nucl. Phys. {\bf B156} (1979)
365.

\bibitem{Bhanot:1979vb}
G.~Bhanot and M.~E. Peskin, \emph{ {Short Distance Analysis for Heavy Quark
  Systems. 2. Applications}}, Nucl. Phys. {\bf B156} (1979)
391.

\bibitem{Klebanov:1997kc}
I.~R. Klebanov, \emph{ {World-volume approach to absorption by non-dilatonic
  branes}}, Nucl. Phys. {\bf B496} (1997) 231--242,
\href{http://www.slac.stanford.edu/spires/find/hep/www?texkey=Klebanov:1997kc}%
{hep-th/9702076}.

\bibitem{Policastro:2002se}
G.~Policastro, D.~T. Son, and A.~O. Starinets, \emph{ {From AdS/CFT
  correspondence to hydrodynamics}}, JHEP {\bf 09} (2002) 043,
\href{http://www.slac.stanford.edu/spires/find/hep/www?texkey=Policastro:2002s%
e}{hep-th/0205052}.

\bibitem{Ghoroku:2004sp}
K.~Ghoroku and M.~Yahiro, \emph{ {Chiral symmetry breaking driven by dilaton}},
  Phys. Lett. {\bf B604} (2004) 235--241,
\href{http://www.slac.stanford.edu/spires/find/hep/www?texkey=Ghoroku:2004sp}{%
hep-th/0408040}.

\bibitem{Gubser:1996de}
S.~S. Gubser, I.~R. Klebanov, and A.~W. Peet, \emph{ {Entropy and Temperature
  of Black 3-Branes}}, Phys. Rev. {\bf D54} (1996) 3915--3919,
\href{http://www.slac.stanford.edu/spires/find/hep/www?texkey=Gubser:1996de}{h%
ep-th/9602135}.

\bibitem{Kovtun:2005ev}
P.~K. Kovtun and A.~O. Starinets, \emph{ {Quasinormal modes and holography}},
  Phys. Rev. {\bf D72} (2005) 086009,
\href{http://www.slac.stanford.edu/spires/find/hep/www?texkey=Kovtun:2005ev}{h%
ep-th/0506184}.

\bibitem{Son:2002sd}
D.~T. Son and A.~O. Starinets, \emph{ {Minkowski-space correlators in AdS/CFT
  correspondence: Recipe and applications}}, JHEP {\bf 09} (2002) 042,
\href{http://www.slac.stanford.edu/spires/find/hep/www?texkey=Son:2002sd}{hep-%
th/0205051}.

\bibitem{Teaney:2006nc}
D.~Teaney, \emph{ {Finite temperature spectral densities of momentum and R-
  charge correlators in N = 4 Yang Mills theory}}, Phys. Rev. {\bf D74} (2006)
  045025,
\href{http://www.slac.stanford.edu/spires/find/hep/www?texkey=Teaney:2006nc}{h%
ep-ph/0602044}.

\bibitem{Kovtun:2006pf}
P.~Kovtun and A.~Starinets, \emph{ {Thermal spectral functions of strongly
  coupled N = 4 supersymmetric Yang-Mills theory}}, Phys. Rev. Lett. {\bf 96}
  (2006) 131601,
\href{http://www.slac.stanford.edu/spires/find/hep/www?texkey=Kovtun:2006pf}{h%
ep-th/0602059}.

\bibitem{Petreczky:2005nh}
P.~Petreczky and D.~Teaney, \emph{ {Heavy quark diffusion from the lattice}},
  Phys. Rev. {\bf D73} (2006) 014508,
\href{http://www.slac.stanford.edu/spires/find/hep/www?texkey=Petreczky:2005nh%
}{hep-ph/0507318}.

\bibitem{Lebellac}
M.~L. Bellac, {\em Thermal Field Theory}.
\newblock Cambridge University Press, 1996.

\bibitem{Peskin}
M.~E. Peskin and D.~V. Schroeder, \emph{ {An Introduction to quantum field
  theory}},. Cambridge, Massachusetts: Perseus-Books, p. 504 problem 15.5
  (1995).

\bibitem{Freedman:1998tz}
D.~Z. Freedman, S.~D. Mathur, A.~Matusis, and L.~Rastelli, \emph{ {Correlation
  functions in the CFT($d$)/AdS($d+1$) correspondence}}, Nucl. Phys. {\bf B546}
  (1999) 96--118,
\href{http://www.slac.stanford.edu/spires/find/hep/www?texkey=Freedman:1998tz}%
{hep-th/9804058}.

\end{thebibliography}\endgroup


\end{document}